\RequirePackage{fix-cm}
\documentclass{svjour3}                     
\smartqed  
\usepackage{graphicx}
\usepackage{nicefrac}
\newcommand{\DOne}{\ensuremath{2s\,^2{\rm{S}}_{\nicefrac{1}{2}} \rightarrow 2p\,^2{\rm{P}}_{\nicefrac{1}{2}}}}
\newcommand{\DTwo}{\ensuremath{2s\,^2{\rm{S}}_{\nicefrac{1}{2}} \rightarrow 2p\,^2{\rm{P}}_{\nicefrac{3}{2}}}}
\newcommand{\DOneTwo}{\ensuremath{2s\;^2\mathrm{S}_{\nicefrac{1}{2}} \rightarrow 2p\;^2\mathrm{P}_{\nicefrac{1}{2},\, \nicefrac{3}{2}}}}

\newcommand\Mark[1]{$^{#1}$}
\newcommand\comma{$^{,}$}

\journalname{}
\begin{document}

\title{Frequency-comb based collinear laser spectroscopy of Be for nuclear structure investigations and many-body QED tests} 
\titlerunning{Frequency-comb based collinear laser spectroscopy}        
\authorrunning{A.~Krieger, W. N\"ortersh\"auser, C. Geppert et al.} 

\author{A.~Krieger\Mark{1}\comma\Mark{2}\comma\Mark{3} \and
        W.~N\"ortersh\"auser\Mark{1}\comma\Mark{2} \and
        Ch.~Geppert\Mark{1}\comma\Mark{2} \and
        K. Blaum\Mark{4}            \and
        M.\,L. Bissell\Mark{5}       \and
        N. Fr\"ommgen\Mark{2}       \and
        M.~Hammen\Mark{2}\comma\Mark{3} \and
        K.~Kreim\Mark{4}            \and
        M.~Kowalska\Mark{6}         \and
        J.~Kr\"amer\Mark{1}\comma\Mark{2}         \and
        R.~Neugart\Mark{2,4}          \and
        G.~Neyens\Mark{5}           \and
        R.~S\'anchez\Mark{2}\comma\Mark{7}        \and
        D.~Tiedemann\Mark{2}               \and
        D.\,T.~Yordanov\Mark{4}\comma\Mark{6}      \and
        M.~Zakova\Mark{2} 
       }
        
\institute{
	\Mark{1} Institut f\"ur Kernphysik, Technische Universit\"at Darmstadt, D-64289 Darmstadt, Germany\\
	\Mark{2} Institut f\"ur Kernchemie, Johannes Gutenberg-Universit\"at Mainz, D-55128 Mainz, Germany\\
	\Mark{3} Helmholtz-Institut Mainz, Johannes Gutenberg-Universit\"at Mainz, D-55128 Mainz, Germany \\
	\Mark{4} Max-Planck-Institut f\"ur Kernphysik, D-69117 Heidelberg, Germany\\
	\Mark{5} Instituut voor Kern- en Stralingsfysica, KU Leuven, B-3001 Leuven, Belgium \\
	\Mark{6} CERN, Physics Department, CH-1211 Geneva 23, Switzerland \\
	\Mark{7} GSI Helmholtzzentrum f\"ur Schwerionenforschung, D-64291 Darmstadt, Germany \\
	W.\,N\"ortersh\"auser \at
            Tel.: +49-6151-16-23575
            \email{wnoertershaeuser@ikp.tu-darmstadt.de}  
}

\maketitle



\begin{abstract}
Absolute transition frequencies of the 
$2s\,^2{\rm{S}}_{\nicefrac{1}{2}}$ $\rightarrow$ $2p\,^2{\rm{P}}_{\nicefrac{1}{2},\nicefrac{3}{2}}$ 
transitions in Be$^+$ were measured with a frequency comb in stable and short-lived isotopes at ISOLDE (CERN) using collinear laser spectroscopy.\\ Quasi-simultaneous measurements in copropagating and counterpropagating geometry were performed to become independent from acceleration voltage determinations for Doppler-shift corrections of the fast ion beam. Isotope shifts and fine structure splittings were obtained from the absolute transition frequencies with accuracies better than 1\,MHz and led to a precise determination of the nuclear charge radii of $^{7,10-12}$Be relative to the stable isotope $^9$Be. Moreover, an accurate determination of the $2p$ fine structure splitting allowed a test of high-precision bound-state QED calculations in the three-electron system. Here, we describe the laser spectroscopic method in detail, including several tests that were carried out to determine or estimate systematic uncertainties. Final values from two experimental runs at ISOLDE are presented and the results are discussed. 

\end{abstract}

\section{Introduction}
\label{intro}
Laser spectroscopy provides a detailed insight into atomic structure including all subtle effects that contribute to the exact energy and the splittings of individual energy levels. Many of these effects are of great relevance in fundamental physics problems, as for example quantum electrodynamics, nuclear structure and weak interaction. Nowadays, laser spectroscopy combined with theoretical calculations is an indispensable tool to explore many-body QED in weak and strong fields and the search for a time or spatial dependence of fundamental constants like the fine structure constant. It provides important information for the analysis of spectra from stars and quasars, for studies of the nuclear structure and for determining the weak charge of a nucleus.  

The technique we present here, has provided new data in two of the mentioned fields, namely the determination of nuclear charge radii and moments of beryllium isotopes \cite{Noe09,Kri12} and the test of many-body bound-state QED calculations in three-electron systems \cite{Noe15}. It is based on collinear laser spectroscopy, a technique that has been contributing to these fields considerably and is one of the workhorses for investigations of nuclear spins and moments, which is witnessed by a long series of review papers \cite{Neu85,Ott89,Bil95,Neu02,Neu06,Che10,Bla13,Cam15} over the last decades. In parallel, it has also been used to investigate the fine structure splittings in helium-like ions as a test of bound-state QED. Such tests were carried out using boron B$^{3+}$ \cite{Din91}, nitrogen N$^{5+}$ \cite{Tho98} and fluorine F$^{7+}$ \cite{Mye99}. In these experiments counter- and copropagating beams have been used to determine absolute frequencies, while for the spectroscopy of short-lived neon isotopes a similar approach was used to calibrate the acceleration voltage of the ions \cite{Gei99,Mar11}. 

For the measurements on beryllium isotopes we have further developed this technique and combined it with a frequency comb to provide high-precision measurements of the transition frequencies. A photon-ion coincidence detection provided the sensitivity required for the detection of the 20-ms isotope $^{12}$Be. These investigations were motivated twofold, by the nuclear structure aspect and the possibility to provide an important benchmark for bound-state QED calculations in three-electron systems.

For nuclear structure physics the nuclear charge radius is an important observable. Its change along a chain of isotopes is extracted with high precision from optical isotope shifts. This provides insight into differences of the radial distribution of protons and the underlying collective effects of soft or rigid deformation or cluster structures, which are often observed for the few-nucleon systems of light nuclei. Only during the last decade new experimental techniques and precise atomic structure calculations for few-electron systems gave access to the determination of charge radii of low-$Z$ nuclei ($Z < 10$) with unprecedented precision. In 2000, first calculations of the mass shift in three-electron systems \cite{Yan00} provided sufficient accuracy to extract the very small finite nuclear size effect from high-precision isotope shift measurements. Since then, calculational precision for three-electron systems has been improved by two orders of magnitude \cite{Yan03,Puc06,Yan08,Yan08b,Noe11}. Pachucki {\it{et al.}} published first results for four-electron systems \cite{Pac04,Puc13} and recently even showed results that pave the way towards boron-like five-electron systems \cite{Puc15b}.

Laser spectroscopy experiments on helium and lithium isotopes were strongly motivated by the existence of so-called halo nuclei. These are nuclear systems with the last neutron(s) being bound by only a few 100\,keV, compared to typical nuclear binding energies of the order of 5--7\,MeV/nucleon. Due to this weak binding, the neutrons are allowed to tunnel far away from the central core, having a large part of their wavefunctions beyond the classical interaction length of the strong force. These nuclei have been a hot topic in nuclear structure research since their discovery in 1983 \cite{Tan85}. Isotope shifts for such systems were measured previously in helium and lithium isotopes. Single atoms of the short-lived two-neutron and four-neutron halo nuclei $^{6,8}$He  were confined in a magneto-optical trap and probed by laser light \cite{Wan04,Mue07}. The lithium isotopes including the two-neutron halo nucleus $^{11}$Li were investigated by applying two-photon resonance ionization spectroscopy \cite{Noe11,Ewa04,San06}. The beryllium isotope chain contains the one-neutron halo nucleus $^{11}$Be and the isotope $^{12}$Be which in the traditional shell model should have a closed neutron shell.

With regard to atomic structure, the vast progress in nonrelativistic few-electron bound-state QED of has opened the possibility of additional tests of many-body QED of in these light systems. The helium fine structure was recently calculated up to the order $m\alpha^7$ and is now one of the most precise QED tests in two-electron systems \cite{Pac10}. The extension of such calculations to three-electron systems proved to be much harder since the extension of the respective computational methods with explicitly correlated functions turned out to be considerably more difficult. Only recently it became possible to perform a complete calculation of $m \alpha^6$ and $m \alpha^7 \ln \alpha$ contributions to the fine structure \cite{Puc14} of a three-electron atom. 
On the experimental side, measurements of the $2p$ fine structure splitting in light three-electron systems are limited in accuracy for isotopes with non-zero nuclear spin due to the unresolved hyperfine structure (hfs) in the $2p_{\nicefrac{3}{2}}$ level. This has been the reason for the fluctuating results on the fine structure splittings in lithium \cite{Noe11,Nob06} being reported for a long time. These turned out to be caused by quantum interference effects in the observation of the unresolved resonance lines \cite{San11}. Once this issue had been resolved experimentally, good agreement with ab initio calculations was obtained \cite{Bro13}. Since relativistic and QED contributions grow in size with increasing $Z$, it became important to study the fine structure splitting also in Be$^+$ to further test bound-state QED. 

Both aspects have been addressed with the technique described in this paper. Besides giving a detailed description of the experiment implemented at ISOLDE (CERN), we will present an overview of the spectroscopic results obtained in two beamtimes (called Run I and Run II). Compared with the techniques used to study helium and lithium isotopes, the collinear approach has the advantage of being more generally applicable and providing high-precision isotope shift data for short-lived isotopes of elements in the so-far inaccessible region $4<Z<10$.

\section{Theory}
\label{sec:Theory}

\begin{table*}
\begin{center}
\caption{Theoretical mass shifts $\delta\nu_{\rm{MS}}^{9,A}$ and field shift factors $F^{9,A}$ for the D1 and D2 transitions \DOneTwo\ in Be$^+$ with respect to $^9$Be$^+$ obtained in two independent calculations \cite{Yan08,Yan08b,Puc08,Puc10} with updated values presented in \cite{Kri12} based on \cite{Dra10,Pac11}. The listed uncertainties are an estimation of unknown higher order terms. The calculations from \cite{Puc10} include another uncertainty that originates from the atomic mass. All values are given in MHz. The deciphered contributions to the mass shift can be found for example in \cite{Puc08}.}
\label{tab:propIS}
\vspace{2mm}
\begin{footnotesize}
\begin{tabular}{r r @{} l r @{} l r @{} l r}
\hline\hline
Isotope & $\delta\nu_{\rm{MS}}^{9,A}$ D1 & & $\delta\nu_{\rm{MS}}^{9,A}$ D2 & &  $F^{9,A}$ & & Reference\\
\hline
$^7$Be$^+$  & -49~225.744(35)&(9) & -49~231.779(35)&(9) &-17.&021(31) & \cite{Puc08,Puc10,Pac11}\\
            & -49~225.779(38)& & -49~231.828(38)& & -16.&912& \cite{Yan08,Yan08b,Dra10} \\
$^{10}$Be$^+$ & 17~310.459(13)&(11) & 17~312.553(13)&(11) &-17.&027(31)&\cite{Puc08,Puc10,Pac11}\\
            & 17~310.442(12)& & 17~312.569(12)& & -16.&912& \cite{Yan08,Yan08b,Dra10}\\
$^{11}$Be$^+$ & 31~560.245(31)&(12) & 31~564.207(31)&(12) &-17.&020(31)&\cite{Puc08,Puc10,Pac11}\\
            & 31~559.990(24)& & 31~563.868(24)& &-16.&912 & \cite{Yan08,Yan08b,Dra10}\\
$^{12}$Be$^+$ & 43~390.180(30)&(180) & 43~395.480(30)&(180) &-17.&022(31)&\cite{Puc08,Puc10,Pac11}\\
            & 43~390.168(39)& & 43~395.499(39)& & -16.&912 & \cite{Yan08,Yan08b,Dra10}\\
\hline\hline\\
\end{tabular}\\
\end{footnotesize}
\end{center}
\end{table*}
It is well known that the isotope shift $\delta\nu^{A,A'}$ between two isotopes $A$ and $A'$ can be separated into the mass shift $\delta\nu_{\rm{MS}}^{A,A'}$ and the field shift $\delta\nu_{\rm{FS}}^{A,A'}$ according to  
\begin{eqnarray}
\label{eq:IS}
\delta \nu_{\mathrm{IS}}^{A,A'}&=&\nu^{A'}-\nu^{A} \\
&=& \underbrace{K_{\mathrm{MS}}\frac{M_{A'}-M_{A}}{M_AM_{A'}}}_{\delta\nu_{\mathrm{MS}}^{A,A'}} +  \underbrace{F^{A,A'} \delta\left\langle r_{\mathrm{c}}^2\right\rangle^{A,A'}}_{\delta\nu_{\mathrm{FS}}^{A,A'}}.
\end{eqnarray}
The mass shift contribution (MS) is related to the center-of-mass motion of the atomic nucleus. For light elements this is the major part of the isotope shift, while the small nuclear volume shift $\delta\nu_{\mathrm{FS}}^{A,A'}$, being typically at the $10^{-5}$ level of the mass shift, contains the information about the change $\delta\left\langle r_{\mathrm{c}}^2\right\rangle$ in the mean square nuclear charge radius. Extraction of nuclear charge radii from experimental isotope shifts in the lightest elements requires accurate mass shift calculations. Semi-empirical techniques that have often been applied for heavier elements to evaluate the atomic parameters $K_{\mathrm{MS}}$ and $F$ are not sufficiently accurate. Only state-of-the-art ab-initio calculations can provide the accurate mass shift and field shift coefficients. Detailed descriptions of these calculations can be found, e.g.,  in \cite{Yan00,Puc06,Noe11,Lu13}. Briefly, the starting point is the non-relativistic Schr\"odinger equation which is solved with high numerical accuracy in the basis of Hylleraas coordinates that explicitly take electron-electron correlations into account. The wavefunctions obtained are then used to calculate relativistic and QED corrections perturbatively as a power series in terms of the fine structure constant $\alpha$. The results for the Be$^+$ isotopes as taken from \cite{Yan08,Yan08b,Dra10,Puc10} are listed in Table~\ref{tab:propIS}. It is worthwhile to note that the calculations performed by two independent groups agree within uncertainties for all isotopes. The only significant difference concerns the case of $^{11}$Be, where the nuclear polarizability correction of 211\,kHz has been calculated and included in \cite{Puc10} but not in \cite{Dra10}. The field shift factor $F^{9,A}$ has been calculated for each isotope individually and is almost constant along the isotopic chain, besides a small difference in the relativistic correction. Using the mass shift values from the table and the measured isotope shifts, the change in the mean square nuclear charge radius can be determined using
\begin{equation}
\delta \langle r_{\mathrm{c}}^2 \rangle^{9,A} = \frac{\delta \nu_{\mathrm{IS}}^{9,A}-\delta \nu_{\mathrm{MS}}^{9,A}}{F^{9,A}}.
\label{eq:changer}
\end{equation}
The absolute charge radius $R_{\mathrm{c}}(A) = \sqrt{\langle r_{\mathrm{c}}^2 \rangle^A}$ of at least one (or more) stable isotope(s) determined by other methods is required to obtain absolute charge radii of the radioactive isotopes. In the case of beryllium the nuclear charge radius of the stable $^9$Be nucleus was determined from elastic electron scattering \cite{Jan72} and thus
\begin{equation}
R_{\mathrm{c}}(^{A}\mathrm{Be}) = \sqrt{R_{\mathrm{c}}^2(^9\mathrm{Be}) + \delta \langle r_{\mathrm{c}}^2 \rangle^{9,A}}.
\label{eq:absoluter}
\end{equation}

\begin{figure*}[bth]
\includegraphics[width=\linewidth]{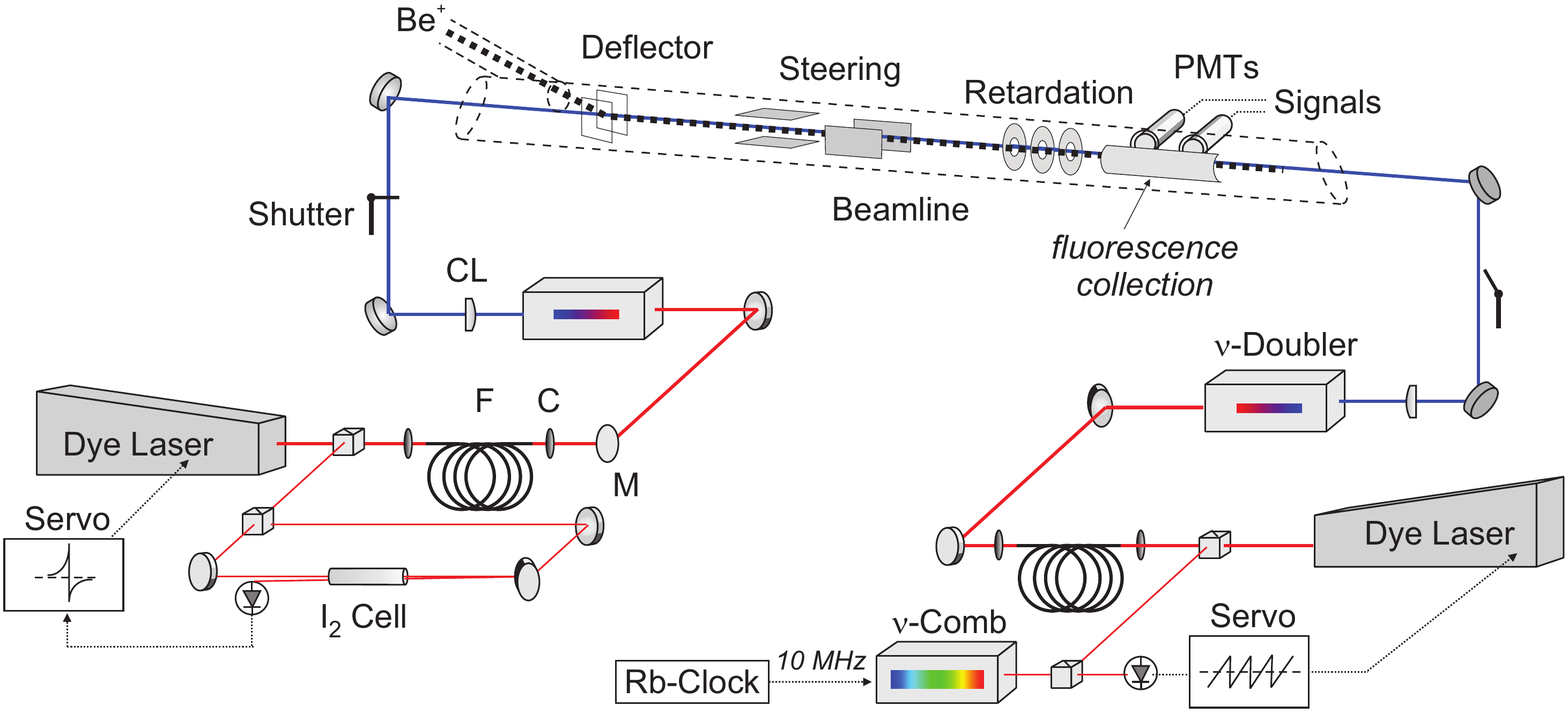}
  \caption{Experimental setup for the beryllium measurements at ISOLDE. Two dye laser systems were used to excite the \DOneTwo\ transitions in Be$^+$. The dye laser for collinear excitation (left) was operated at a fundamental wavelength of 624\,nm and stabilized to a hyperfine transition of molecular iodine. The output beam is frequency-doubled to 312\,nm and guided into the beam line. The other laser (right) is locked to a frequency comb. After frequency-doubling to 314\,nm the UV laser beam is anticollinearly superposed with the ion beam. The resonance fluorescence is detected by a pair of photomultipliers. A photon-ion coincidence detection unit increases the detection efficiency if the ion beam rate is low (not shown, see Fig.\,\ref{fig:COLLAPSel}).}	
	\label{fig:COLLAPSlasersetup}
\end{figure*}
 
The many-electron Dirac equation poses some difficulties for the inclusion of relativistic effects and correlations between electrons in atomic systems. According to QED the equation has to include multiple electron-positron pairs, which leads to  numerical instabilities. This problem limited the relativistic calculation of the lithium
$2p\,^2\mathrm{P}_{\nicefrac{1}{2}}$ -- $2p\,^2\mathrm{P}_{\nicefrac{3}{2}}$ splitting to one significant digit \cite{Der08}. Forty years after first numerical calculations using explicitly correlated basis sets with Hylleraas and
Gaussian functions for two electrons \cite{Dou74}, Puchalski and Pachucki extended such calculations to three-electron systems \cite{Puc14}. Nonrelativistic QED can perturbatively account for relativistic, retardation, electron self-inter\-action, and vacuum polarization contributions by an expansion of the level energy in
powers of the fine structure constant $\alpha$
\begin{equation}
E = m\alpha^2 E^{(2)}  + m\alpha^4 E^{(4)} + m\alpha^5 E^{(5)} + m\alpha^6 E^{(6)} + ...  \, , 
\end{equation}
where the expansion coefficients $E^{(i)}$ may include powers of $\ln \alpha$. In this expansion, the fine structure arises at the order of $m\alpha^4$, together with the nuclear recoil term, which in this order is comparable in size to $m\alpha^6$ contributions, but of opposite sign. For all details of the calculations and the individual contributions we refer to \cite{Puc15}.
In the splitting isotope shift (SIS), i.e. the difference in fine structure splitting between isotopes, all mass-independent terms cancel and only the mass-dependent terms remain, which can be calculated with very high accuracy. 

The SIS therefore provides a valuable consistency check of the experimental results \cite{Yan02}. For isotopes with nuclear spin, hyperfine-induced fine-structure mixing can lead to an additional level shift that also contributes to the SIS. This in combination with the unresolved hyperfine splittings in the $2p\,^2\mathrm{P}_{\nicefrac{3}{2}}$ level in light three-electron systems makes even-even isotopes with nuclear spin $I=0$ an exceptionally suitable case to perform tests of the calculations. While there is no such isotope for lithium, the beryllium chain with $^{10}$Be and $^{12}$Be includes two spinless isotopes that are accessible to the measurement. 

\section{Experimental Setup}

A schematic overview of the experimental setup applied for collinear laser spectroscopy on Be$^+$ ions in the \DOne\ (D1) and \DTwo\ (D2) transitions is shown in Fig.~\ref{fig:COLLAPSlasersetup}. A mass-separated ion beam of a stable or a radioactive beryllium isotope at an energy up to 60~keV was transported to the laser beam line. Two frequency-stabilized dye laser systems delivered UV beams that superposed the beryllium ion beam in opposite directions. The resonance fluorescence photons were detected via photomultipliers. The resonance condition was established by tuning the Doppler-shifted frequency with an electrical potential applied to the fluorescence detection chamber. 
The individual parts of the experimental setup as well as the scanning procedure are described in detail in the following subsections.
 
\subsection{Production of Radioactive Beryllium Isotopes}
The stable and radioactive beryllium isotopes were produced at the on-line isotope separator facility ISOLDE at CERN. High-energy (1.4~GeV) protons from the PS-Booster synchrotron impinge on a uranium carbide target. The atoms are photo-ionized using the resonance ionization laser ion source RILIS \cite{Fed08}. Resonant excitation at 234.9~nm from the atomic ground state in the $2s^2$ $^{1}{\rm{S}}_0 \rightarrow 2s 2p$ $^{1}{\rm{P}}_1$ transition, followed by excitation at 297.3~nm to the auto-ionizing $2p^2$ $^{1}{\rm{S}}_0$ level was employed to ionize the Be atoms which have the rather large ionization potential of 9.4~eV.

Table~\ref{tab:isoldeyield} lists the ion beam intensities decreasing from $^{7}$Be to $^{12}$Be by seven orders of magnitude \cite{Koe98}. 
In the final stage of our experiment an upgraded solid-state pump laser system \cite{Fed08} was used. This gave a $^{11}$Be yield of up to $2.7 \cdot 10^7$ ions/s, about 4 times larger than reported previously.

The yields are sufficient to perform collinear laser spectroscopy on $^{7-11}$Be solely based on a standard fluorescence detection system. However, for a beam intensity of less than 10$^4$~ions/s, i.e. for measurements on $^{12}$Be, the sensitivity had to be enhanced. This was achieved by detecting ion-photon coincidences and thus rejecting the stray light background which usually determines the sensitivity limit. Coincidence detection requires an isobarically clean ion beam. For that reason the pulse structure and possible contamination of the beam was investigated and optimized for $^{12}$Be.

\begin{table}
\begin{center}
\caption{Nuclear properties and production rates of the beryllium isotopes at the ISOLDE facility at CERN. The table includes the half-life ($T_{1/2}$), nuclear spin $I$, magnetic dipole moment ($\mu_I$) in nuclear magnetons ($\mu_{\mathrm{N}}$) and the yields using a 1.4-GeV proton beam from the PS booster and RILIS for ionization \cite{Iso13}.}
\label{tab:isoldeyield}\vspace{2mm}
\begin{tabular}{l l c r l c}
\hline\hline
 &  T$_{1/2}$  & I & \multicolumn{1}{c}{$\mu_{I}/\mu_{\mathrm{N}}$}&  $({\rm{ions}}/\mu {\rm{C}})$ \\
\hline 
~$^{7}$Be  &  53 d      & 3/2   & --1.39928(2)\,\cite{Oka08}  & $1.4 \cdot 10^{10}$ \\
 
~$^{9}$Be  & stable     & 3/2   & --1.177432(3)\,\cite{Win83} &  \\

$^{10}$Be  & $1.6\cdot10^6$\,a&0& ~~~--~~~                    & $6.0 \cdot 10^{9}$\\

$^{11}$Be  & 13.8 s     & 1/2   & -1.6813(5)~~\cite{Noe09}    & $7.0 \cdot 10^{6}$\\

$^{12}$Be  & 23.8 ms    & 0     &  ~~~--~~~                   & $1.5 \cdot 10^{3}$\\

$^{14}$Be  & 4.35 ms    & 0     &   ~~~--~~~                  & $4.0 \cdot 10^{0}$\\
\hline\hline \\
\end{tabular}
\end{center}
\end{table}

\subsection{Beryllium Ion Beam Structure} \label{sec:ionprop}
\begin{figure}
\begin{center}
\includegraphics[width=0.9\linewidth]{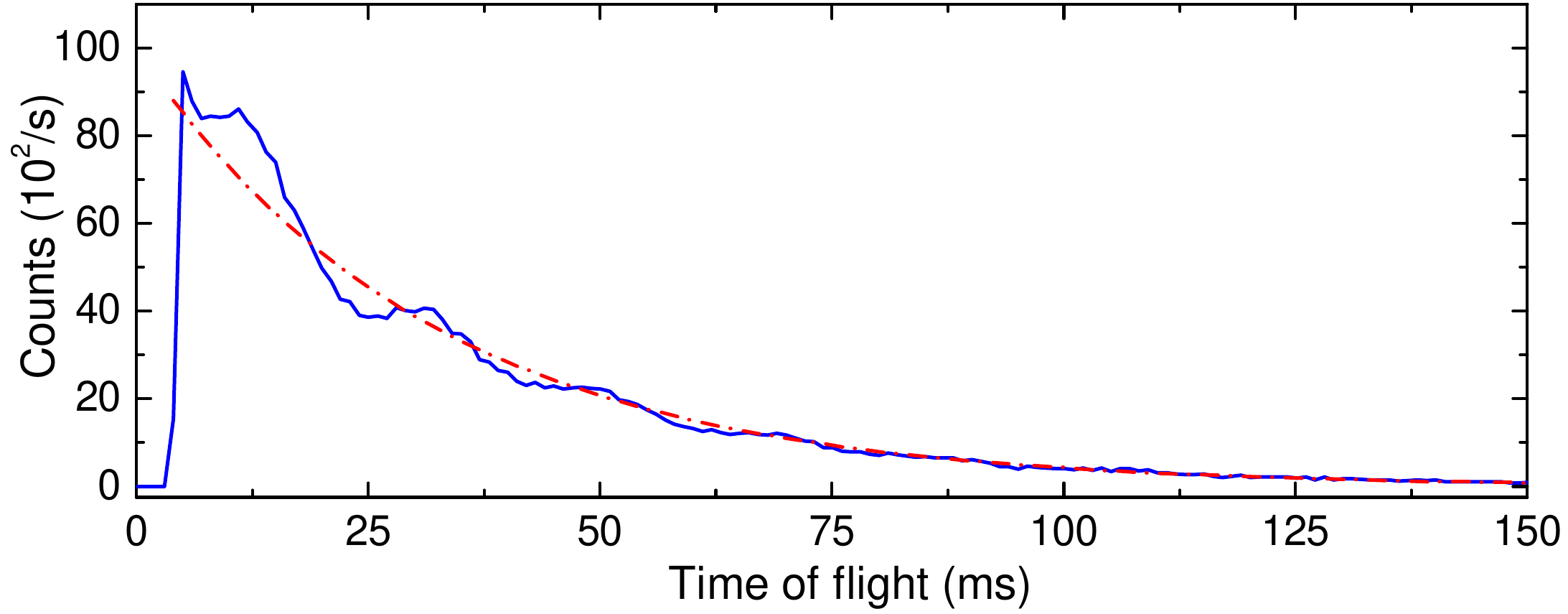}
\end{center}
  \caption{Release of beryllium ions (solid blue line) from ISOLDE as a function of time after the proton pulse hit the target container, measured with a secondary electron multiplier at the end of the COLLAPS beam line. The release curve, integrated over 100 proton pulses with a resolution
of 0.2\,ms/channel, is modelled with an exponential decay curve (dash-dotted red line).(Color online)}	
	\label{fig:release}
\end{figure}

During our experiment at ISOLDE, pulses of $3 \cdot 10^{13}$ protons impinged on a UC$_x$ target typically every 4\,s. The release of resonantly ionized $^{12}$Be was tracked using a secondary electron multiplier installed at the end of the laser spectroscopy beam line. 
The proton pulses triggered a multichannel analyzer that recorded the ion events as a function of time. Figure\,\ref{fig:release} shows such a release curve summed over 100 proton pulses with a resolution of 0.2\,ms/channel. The integral corresponds to a release of 12\,000\,ions per proton pulse. This is almost a factor of 10 more than listed in the yield table (Tab.\,\ref{tab:isoldeyield}). During the measurements on $^{12}$Be the typical ion yield was about 8\,000\,ions/pulse. 

The release curve of Fig.\,\ref{fig:release} demonstrates a characteristic feature of the ISOLDE HV supply: First ions are detected about 2-3\,ms after the proton pulse hit the target. This delay is determined by the recovery time of the high voltage, which is pulsed down right before a proton pulse hits the target, in order to reduce the current load from ionized air \cite{Fia92}. After an initial steep rise the release curve follows essentially the exponential decay of $^{12}$Be. The extracted half-life of $T_{1/2} \approx 21.9(8)$\,ms agrees well with the literature value of 21.50(4)\,ms \cite{Aud97}. The single exponential does not exhibit any significant offset. This demonstrates that practically no beam contamination from the isobar $^{12}$C$^+$ is present and $^{12}$B$^+$ (having a similar lifetime as $^{12}$Be) is also not expected due to the relatively high ionization potential. This situation is prerequisite for the application of a photon-ion coincidence technique, which otherwise would suffer from random coincidences between scattered laser light and isobaric ions. With the rapid decay of $^{12}$Be the fluorescence detection can be limited to about 100~ms after the proton pulses. 

\subsection{Experimental Beam Line} \label{sec:COLLAPSsetup}
The COLLAPS collinear spectroscopy beam line at the ISOLDE facility was commissioned in the early eighties \cite{Neu81,Buc82,Mue83} and has been improved continuously \cite{Neu06,Gei99,Neu86,Gei05,Ney05,Kow05} with the objective of widening the range of accessible elements and isotopes. An important aspect was the development of highly sensitive alternatives to the traditional flourescence photon detection technique. 
\begin{figure*}
\begin{center}
\includegraphics[width=0.85\linewidth,clip=true]{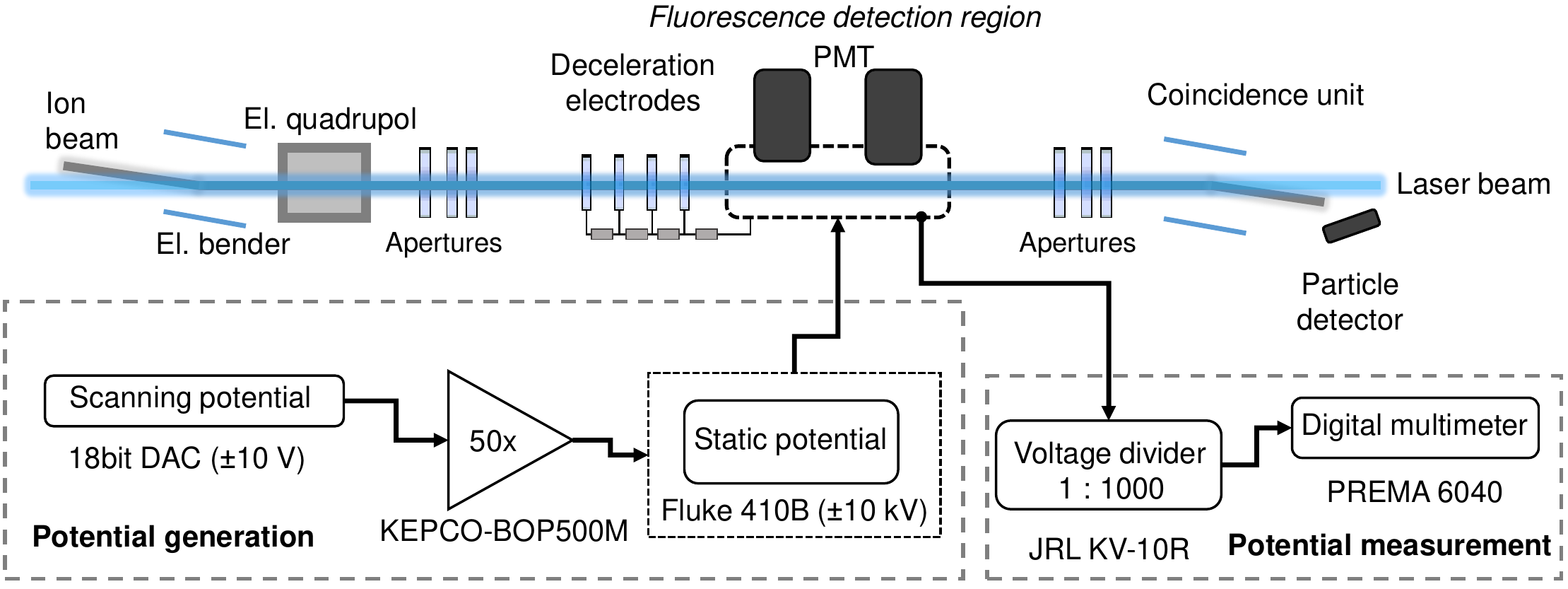}
\end{center}
  \caption{Schematic view of the COLLAPS beam line at ISOLDE/CERN and corresponding high-voltage circuits: A mass-separated ion beam is directed along the axis of the vacuum beam line with the help of an electrostatic deflector. An electric dipole and quadrupole collimates and steers the beam through the apparatus. The fluorescence detection region is a cage floated on a variable potential against ground to enable Doppler-tuning. At the end of the beam line a photon-ion coincidence detection chamber is installed, whereby a secondary electron multiplier is used to count the ions. The generation and measurement of the high-voltage potential is explained in the text.}	
	\label{fig:COLLAPSel}
\end{figure*}
For conventional collinear spectroscopy the ions are accelerated to a beam energy of typically 50\,keV, with the corresponding positive potential applied to the ion source, while the mass separator and the experimental beam line are on ground potential. The ion beam is merged with a laser beam by a pair of deflector plates as shown in Figs.\,\ref{fig:COLLAPSlasersetup} and \ref{fig:COLLAPSel}. 

A quadrupole triplet collimates the ion beam, matching it to the laser beam profile, and a second set of deflector plates aligns it with the laser beam axis which is defined by two apertures at a distance of about 2\,m. Two UV-sensitive photomultiplier tubes with 45\,mm active aperture and 15\,\% quantum efficiency at 313\,nm are used for fluorescence detection. The light collection system consists of two fused-silica lenses of 75\,mm diameter and a cylindrical mirror opposite to them.
Stray laser light is suppressed by sets of apertures with diameters decreasing with distance from the optical detection region and Brewster-angle quartz windows at both ends.

Collinear laser spectroscopy is usually performed with the laser running at a fixed frequency, while the absorption frequency of the ions is tuned by changing their velocity (Doppler-tuning). This means that a variable electrical potential has to be applied to the interaction region. For applying post-acceleration/deceleration voltages up to 10\,kV a set of four electrodes provides a smoothly variable potential along the beam axis. In order to avoid optical pumping into dark states, the final ion velocity is reached just in front of the detection region by applying a small fixed offset voltage between the last electrode and the detection chamber. 
The lower part of Fig.~\ref{fig:COLLAPSel} illustrates the generation of the voltage between the detection region and ground as a combination of a static high voltage in the range of $\pm 10$~kV and a scanning voltage of $\pm 500$\,V. The latter is created by amplification ($\times 50$) of the $\pm$ 10\,V dc output of an 18-bit DAC controlled by the measuring computer. 
This voltage defines the floating offset potential of a stabilized $\pm 10$\,kV power supply. The combination of power supplies makes it possible to perform measurements on a series of isotopes with different Doppler shifts and for each of them scan small frequency ranges covering the hyperfine structure with high resolution. 

Both the static high voltage and the scanning voltage are measured with a high-precision 1:1000 voltage divider and a digital voltmeter. A comparison of the measured voltages with those obtained using a 
precision voltage divider calibrated at PTB (Braunschweig, Germany) 
\cite{Thu09} has demonstrated an uncertainty of $\nicefrac{\Delta U}{U} < 3 \cdot 10^{-5}$ which corresponds to about 0.3\,V at a maximum 
voltage of 10\,kV applied to the excitation region. Still, the knowledge 
of the ion beam velocity is limited by the uncertainty of the ion source 
potential which is determined by the main acceleration voltage power 
supply. Also the specified accuracy $\nicefrac{\Delta U}{U} < 1 \cdot 
10^{-4}$ of the voltage measurement on the operational voltage of 
60\,kV was verified by calibration with the precision voltage divider \cite{Kri11}.
It translates to less than 6\,eV uncertainty in the ion beam energy.

As in the laboratory frame the transition frequency of the ions in collinear geometry scales as
\begin{equation}
	\nu_{c} = \nu_0 \, \gamma \, \left( 1 + \beta \right)
	\label{eq:vnu}
\end{equation}
where the dimensionless ion velocity is $\beta = v/c$, the relativistic factor $\gamma = \sqrt{1/(1-\beta^2)}$ and the transition frequency $\nu_0$, any uncertainty in $\beta$ arising from the ion source potential results in an uncertainty of measured absolute transition frequencies or isotope shifts, especially for light ions. In the particular case of beryllium a deviation of 6\,V from the measured voltage result in an artificial isotope shift of $\delta \nu^{9,11} \left(^{9} \textrm{Be},^{11}\textrm{Be} \right) = 18$\,MHz. 
To overcome these limitations, we have introduced the (quasi-)simultaneous excitation by a collinear and an anticollinear laser beam. The method is based on the fact that in this geometry the measured resonance frequencies, $\nu_{c} = \nu_0 \gamma (1+\beta)$ for collinear and $\nu_a = \nu_0 \gamma (1-\beta)$ for anticollinear excitation are simply related to the rest frame frequency $\nu_0$ by
\begin{equation}
	\nu_{c} \cdot \nu_{a} = \nu^2_0 \gamma^2 \cdot \left( 1 + \beta \right)\left( 1 - \beta \right) =  \nu^2_0.
	\label{eq:vnull}
\end{equation}
This provides a method to determine the transition frequency independently of the knowledge of the ion beam energy which depends on assumptions about the ion source potential and on measured voltages. However, in contrast to conventional collinear laser spectroscopy, this approach requires two laser systems instead of one and, additionally, the capability to determine the absolute laser frequencies with an accuracy better than $10^{-9}$. Similar approaches were proposed and demonstrated for the measurement of absolute transition frequencies \cite{Pou88} and used for, e.g., precision spectroscopy in the fine structure of helium-like Li$^+$, yielding an accurate value of the Lamb shift \cite{Rii94}. Here we have developed a procedure which is widely applicable in cases where high precision is required for the spectroscopy of unstable isotopes. 

\subsection{Setup and Specification of the Frequency-Comb-Based Laser System} \label{sec:COLLAPSlaser}
The transition wavelength  of the \DOneTwo\ transitions in Be$^+$ is about 313\,nm corresponding to an energy splitting of $\approx 4$\,eV. The laser system installed at COLLAPS is schematically shown in Fig.\,\ref{fig:COLLAPSlasersetup}. For anticollinear excitation a frequency doubled Nd$:$YVO$_4$ laser (Verdi V18) was operated at 9\,W to pump a Coherent 699-21 dye laser. Using a dye solution of Sulforhodamine B in ethylene glycol, a typical output power of 700\,mW was achieved at the wavelength of 628\,nm. Another dye laser, Sirah MATISSE DS, was installed for collinear excitation and operated with a dye solution of DCM in 2-phenoxy-ethanol. With the 8-W pump beam from a Verdi V8, about 1.2\,W were achieved at the fundamental wavelength of 624\,nm. Each laser beam was then coupled into a 25-m long photonic crystal fiber (LMA-20) to transport the laser light to one of the second harmonic generators installed in the ISOLDE hall. A two-mirror delta cavity (Spectra Physics Wavetrain) and a four-mirror bow-tie cavity (Tekhnoscan FD-SF-07) were located nearby the COLLAPS beam line. A Brewster-cut and an anti-reflection coated BBO crystal, respectively, converted the laser beams of 628\,nm and 624\,nm into their second harmonics at 314\,nm and 312\,nm, in both cases with an output of more than 10\,mW. The elliptical UV beams were reshaped to circular beams with diameters of 3-4\,mm to match the transversal profile of the ion beam and finally attenuated to powers below 5\,mW. Two remote-controlled fast beam shutters blocked alternatively the collinear or the anticollinear laser beam. This enabled us to perform scans of 3-30\,s duration in collinear or anti-collinear configuration in a fast sequence.

\begin{figure}
\includegraphics[width=\linewidth,clip=true]{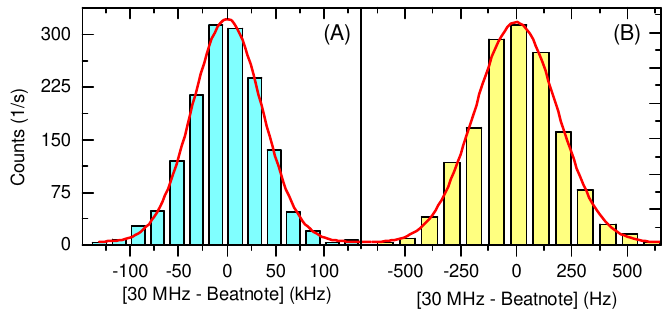}
  \caption{Beat frequency histograms of the frequency-stabilized dye lasers. The beat was averaged for 1\,s and the distribution of the beat frequencies over a period of 2 hours is shown in the histogram. Graph (A) shows the histogram of the MATISSE DS laser stabilized to a hyperfine transition of molecular iodine. The corresponding uncertainty was estimated as the FWHM of a Gaussian fit (red) of about 75\,kHz. The results of the frequency-comb stabilized Coherent 699-21 dye laser is depicted in Graph (B). In this case the FWHM is approximately 400\,Hz.}
\label{fig:coherentdw}
\end{figure}

The backbone of the laser system was the precise frequency stabilization and frequency measurement required for the application of Eq.\,(\ref{eq:vnull}). In practice, the transition rest-frame frequency $\nu_0$ depends on the absolute output frequencies of both dye lasers which have to be known with a relative accuracy better than $\Delta\nu/\nu \leq 10^{-9}$ to yield the isotope shifts with an accuracy better than $10^{-5}$. 
Therefore, a Menlo Systems frequency comb (FC 1500) with a repetition frequency of 100\,MHz was employed. A Stanford Research rubidium clock (PRS10) provided the 10-MHz reference for the stabilization of the carrier-envelope-offset (CEO) and the repetition frequency. This  clock was long-term stabilized using a GPS receiver tracking the 1-pps signal.

The MATISSE dye laser for collinear excitation was stabilized to its internal reference cavity for short-term stability. In this case frequency drifts were further reduced by locking the laser to a hyperfine transition in molecular iodine using frequency-modulated saturation spectroscopy. In total 12 hyperfine transitions of $^{127}$I$_2$ match the desired Doppler-shifted frequencies for a wide range of acceleration voltages between 30--60\,kV. The demodulated dispersion signal from the phase-sensitive detection was fed into a 16-bit National Instruments DAQ card (NI-DAQ 6221) and further processed with the MATISSE control software to provide a counter-drift for the MATISSE reference cavity. In regular time intervals the laser frequency was measured with the frequency comb and recorded for a few 100\,s to ensure the stability of the locking point and to provide the absolute frequency for the application of Eq.\,(\ref{eq:vnull}). A histogram of 1-s averaged beat signals measured over 2 hours is depicted in Fig.~\ref{fig:coherentdw}(A). It exhibits a Gaussian distribution with standard deviation of about 75~kHz.  
The frequency of the Matisse laser stabilized to the various iodine lines was repeatedly measured during the beamtimes. The averaged results are listed in Table~\ref{tab:AF1} and compared with the calculated frequencies from \cite{Kno04}. Reasonable agreement is obtained in all cases.
\begin{table*}
\begin{center}
	\caption{Frequencies of the $a_1$ hyperfine component in various transitions of iodine $^{127}$I$_2$ utilized and determined during the experiment. The total uncertainty of the experimental values is about 190\,kHz. The calculated frequencies (theory) are afflicted with an uncertainty of approximately 3\,MHz \cite{Kno04}.}
	\label{tab:AF1}
\begin{tabular}{llll}
\hline\hline
 HFS a$_1$  & Frequency         & Frequency         & Deviation\\
 transition &(theory) (MHz)     & (experiment) (MHz) & (MHz)\\
\hline
R(62)(8-3)  & 479~804~354.67    & 479~804~355.09    &--0.42\\
R(70)(10-4) & 479~823~072.75    & 479~823~072.58    &~0.17\\
P(64)(10-4) & 479~835~709.4     & 479~835~708.96    &~0.44\\
R(60)(8-3)  & 479~870~011.92	& 479~870~012.20    &--0.28\\
R(58)(8-3)  & 479~933~416.07	& 479~933~416.36    &--0.29\\
R(56)(8-3)  & 479~994~568.08    & 479~994~568.11    &--0.03\\
R(54)(8-3)  & 480~053~468.95	& 480~053~469.06    &--0.11\\
R(52)(8-3)  & 480~110~119.57	& 480~110~119.59    &--0.02\\
R(50)(8-3)  & 480~266~578.9     & 480~266~578.59    &~0.31\\
R(48)(8-3)  & 480~314~237.19	& 480~314~236.83    &~0.36\\
R(42)(8-3)  & 480~359~649.42	& 480~359~649.13    &~0.29\\
R(40)(8-3)  & 480~402~816.3	    & 480~402~815.78    &~0.52\\
\hline\hline\\
\end{tabular}\\
\end{center}
\end{table*}

The Coherent 699 dye laser for anticollinear excitation was internally stabilized to its own reference cavity of Fabry-Perot type, long-term frequency drifts were corrected by an additional stabilization to the frequency comb. Therefore the beat signal between the dye laser and the nearest frequency comb mode was detected on a fast photo diode and fed into the Menlo Systems phase comparator DXD 100.  A low-noise PI regulator (PIC 210) processed the signal from the phase comparator and provided a servo-voltage to counteract all frequency excursions of the dye laser by correcting the length of the reference cavity.
As a measure of the long-term stability a beat signal with the frequency comb was detected. The result is shown in Fig.\,\ref{fig:coherentdw}(B). The standard deviation over 2 hours measuring time and 1-s averaging time is about 400~Hz.

 \section{Measurement Procedure} \label{sec:procedureberyllium}

 The ion beam acceleration voltage at the ISOLDE front end was fixed to 
 40\,kV.
 A suitable iodine line is chosen such that the isotope under investigation 
 can be recorded by applying an offset voltage in the available range of 
 $U_{{\rm{Offset}}}=\pm 10$\,kV at the fluorescence detection region. For 
example, when choosing the $a_1$ hyperfine component in the transition 
R(56)(8-3) as a reference, the isotopes $^{9-12}$Be can be addressed. 
The scan voltage range $U_{\rm{scan}}$ of up to $\pm 500$\,V is then 
adjusted to cover the full hyperfine structure in the collinear 
direction and the expected position of the center of gravity is 
calculated. The required offset voltage as well as the scan voltage 
range was estimated based on previous measurements of the beryllium 
absolute transition frequency \cite{Bol85} and nuclear moments 
\cite{Gei99} in combination with the precisely calculated mass shift 
\cite{Dra10,Puc10}. Once the resonance position of $^9$Be was found, the
laser frequencies $\nu_c$ and $\nu_a$ could be predicted for all 
radioactive beryllium isotopes with an accuracy of a few MHz, 
which is the size of the expected field shift contribution.
This knowledge allowed us to calculate the required frequency of the 
second dye laser to simultaneously cover the full hyperfine structure in
anticollinear geometry within the same Doppler tuning voltage range and 
even to ensure that the centers of gravity of both hyperfine spectra
practically coincide within a few 100\,mV, corresponding to the size 
of the field shift contribution. This is only possible because this dye 
laser is locked to the frequency comb and thus can be stabilized at any 
arbitrarily chosen frequency.

Fast laser beam shutters placed in front of the Brewster windows of the 
apparatus were controlled by the data acquisition software in order to 
allow only one of the two laser beams to enter. 
For the isotopes with half-lives longer than the typical 4-s repetition time of proton pulses, fast scans of the Doppler-tuning voltage $U_{{\rm{scan}}}$ were performed with alternating laser beams. The scanning range was chosen depending on the hyperfine splitting of the respective isotope and spectra were taken in 200 channels for $^{10}$Be and up to 800 channels for the odd-$A$ isotopes $^{7,9,11}$Be. The common dwell time was 22\,ms per voltage step. 
Depending on the ion beam intensity, a single spectrum is the 
sum of 50--800 individual scans for each direction.
This procedure was applied using about 3--4 different iodine lines for 
each isotope.

Because of the short 21.5-ms half-life of $^{12}$Be, photon counts had to be accumulated for typically 60\,ms after each proton pulse. The laser shutters for collinear and anticollinear beams were switched between consecutive pulses and the voltage steps were triggered by every second pulse. Given the extremely low ion beam intensity, the single-line spectrum of $^{12}$Be was taken in only 20 channels with a total measuring time of about 8 hours, corresponding to 200 scans.

 \begin{figure}[bth]
 \begin{center}
 {\includegraphics[width=\linewidth]{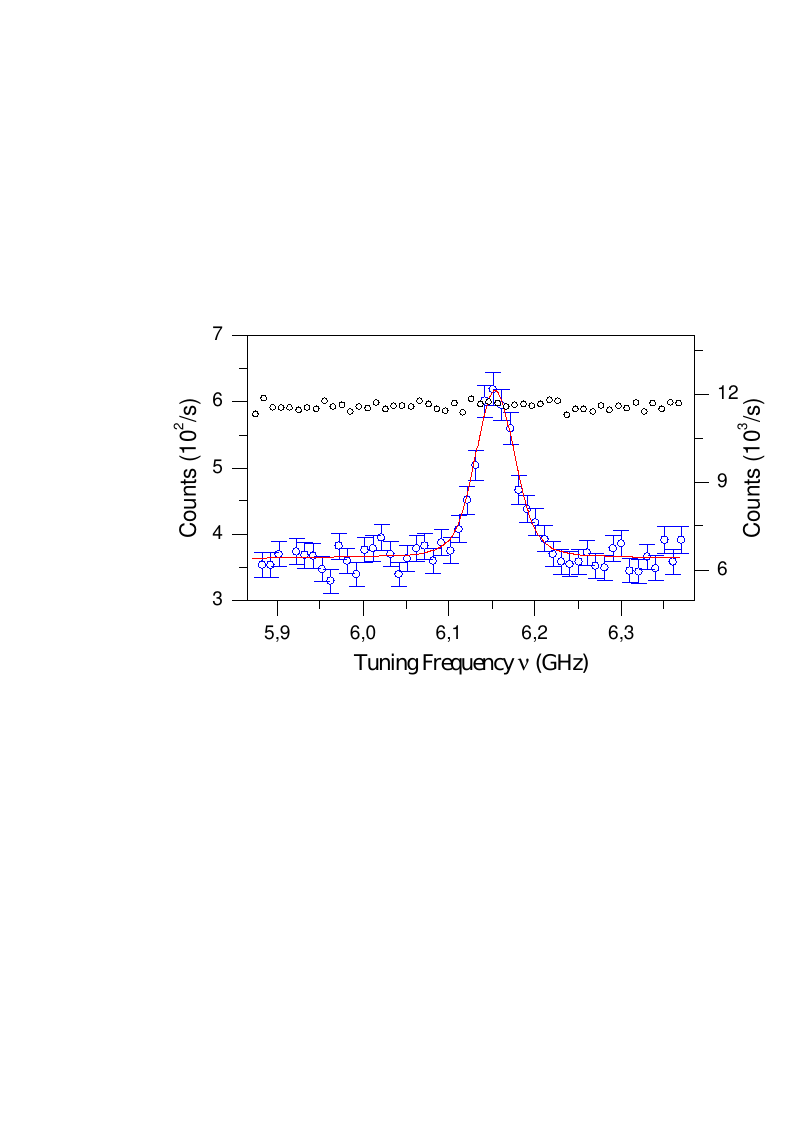}}
    \caption{Comparison between conventional optical fluorescence 
 detection (upper trace, right $y$-axis) and photon ion coincidence 
 detection (lower trace, left $y$-axis) at an ion beam rate of 
 30\,000\,$^{10}$Be$^+$ ions/s. The optical spectrum (black circles) in the 
 conventional detection is covered by stray light of the laser beam. In 
 the photon-ion coincidence spectrum a clear resonance (blue circles) is 
 observed, fitted with a Voigt profile.}
 \label{Fig:bepicoincidence}
 \end{center}
 \end{figure}

Detection of the weak $^{12}$Be signals required the additional rejection of background from scattered laser light reaching the photomultiplier tubes. This was achieved by implementing
a photon-ion coincidence: 
 Photomultiplier signals were accepted only if an ion was simultaneously 
 traversing the detection unit. Downstream of the photon detection chamber the ions were deflected onto the cathode of a secondary 
 electron multiplier (SEM) installed off-axis. Discriminated pulses from the photomultipliers were delayed by the appropriate time of flight (TOF) (3--4\,$\mu$s) of the ion to the SEM. 
 To avoid electronic dead times, the delay was realized logically in a 
 first-in first-out (FIFO) queue structure on a field-programmable gate 
 array (FPGA) with a resolution of 10\,ns, based on the FPGA's internal 
 clock. Signals leaving the queue were transformed back into a TTL pulse
 and fed together with the SEM pulses into a standard coincidence unit.
 The photon-ion coincidence detection was optimized using a $^{10}$Be$^+$
 ion beam,  attenuated to about 30.000\,ions/s by detuning the RILIS 
 laser. The time of flight for $^9$Be was determined with a multi-channel 
 analyzer and the respective TOF for $^{10}$Be was calculated.
 Figure~\ref{Fig:bepicoincidence} shows a comparison between the 
 conventional ungated spectrum (grey circles) and the optical spectrum 
 detected in delayed coincidence (blue circles). The resonance peak is 
 only visible in the gated spectrum. The background induced by laser 
 stray-light was reduced by a factor of 35. However, it must be noted 
 that this reduction factor strongly increases with a reduction of the 
 ion beam rate.
 
  \section{Analysis and Results}
 Two beam times were performed to investigate first the isotopes 
 $^{7-11}$Be (Run I) \cite{Noe09} and then concentrate on 
 $^{12}$Be (Run II) \cite{Kri12} after installing the 
 ion-photon coincidence setup. The stable isotope $^9$Be and the 
 even-even isotope $^{10}$Be were used as reference isotopes, 
 respectively. We concentrate here on results and procedures from Run II 
 and provide differences to Run I only when it is of importance.

 \subsection{Line Shape Studies on $^9$Be and $^{10}$Be} \label{sec:refscans}

Resonance spectra of $^9$Be in the \DOne tran\-sition are shown in the 
 upper trace of Fig.~\ref{Fig:9benew} taken in collinear (left) and 
 anticollinear geometry (right) as a function of the Doppler-tuned laser 
 frequency. Each spectrum is the sum of 20 individual scans. To avoid 
 saturation broadening, the laser beam was attenuated to 3\,mW and 
 collimated to a beam diameter of about 3--4\,mm to match approximately 
 the size of the ion beam. Similar spectra for $^{10}$Be are shown 
in the lower traces. These are the integral of 50 single scans
 at an ion beam current of 10\,pA. A best fit of the resonance was obtained for a
double Voigt profile with a full 
 width half maximum (FWHM) of 40~MHz. It 
 becomes apparent that each peak in the hyperfine structure is actually 
 a composition of two components: A satellite peak, with a 
 typical intensity of less than 5\% of the corresponding main peak, 
 appears on the low-energy tail in each spectrum independent of the 
 direction of excitation. It is induced by a class of ions which have 
 lost some of their kinetic energy. The loss is  
 almost exactly 4\,eV and can be explained by inelastic collisions with residual gas atoms that lead to excitations into the $2p$ states. The energy required for this excitation is taken from the kinetic energy of the ion
and is lost when the excited ion decays to the ground state by 
 emitting a photon. The overall line shape is 
 reasonably well fitted using a Voigt doublet and only small structures remain in the residua, 
 depicted below each spectrum. The remaining small asymmetry seen in this structure is similar for the different peaks. It is an asset of the technique that asymmetries in the collinear and 
 the anticollinear spectra shift the peak center to slightly lower and 
 slightly larger frequencies, respectively. Hence, these shifts 
 largely cancel when calculating the rest frame frequency.

 \subsection{Hyperfine Fitting Procedure}
 \label{sec:lineshapefitting}
 Fitting was performed as follows: Each voltage information was converted 
 into the corresponding Doppler-shifted laser frequency to account for 
 the small nonlinearities in the voltage-frequency relation. Hyperfine 
 peak positions relative to the center of gravity $\nu_{\mathrm{cg}}$ 
 were calculated based on the Casimir formula:

\begin{figure}
 \begin{center}
 \includegraphics[width=\linewidth]{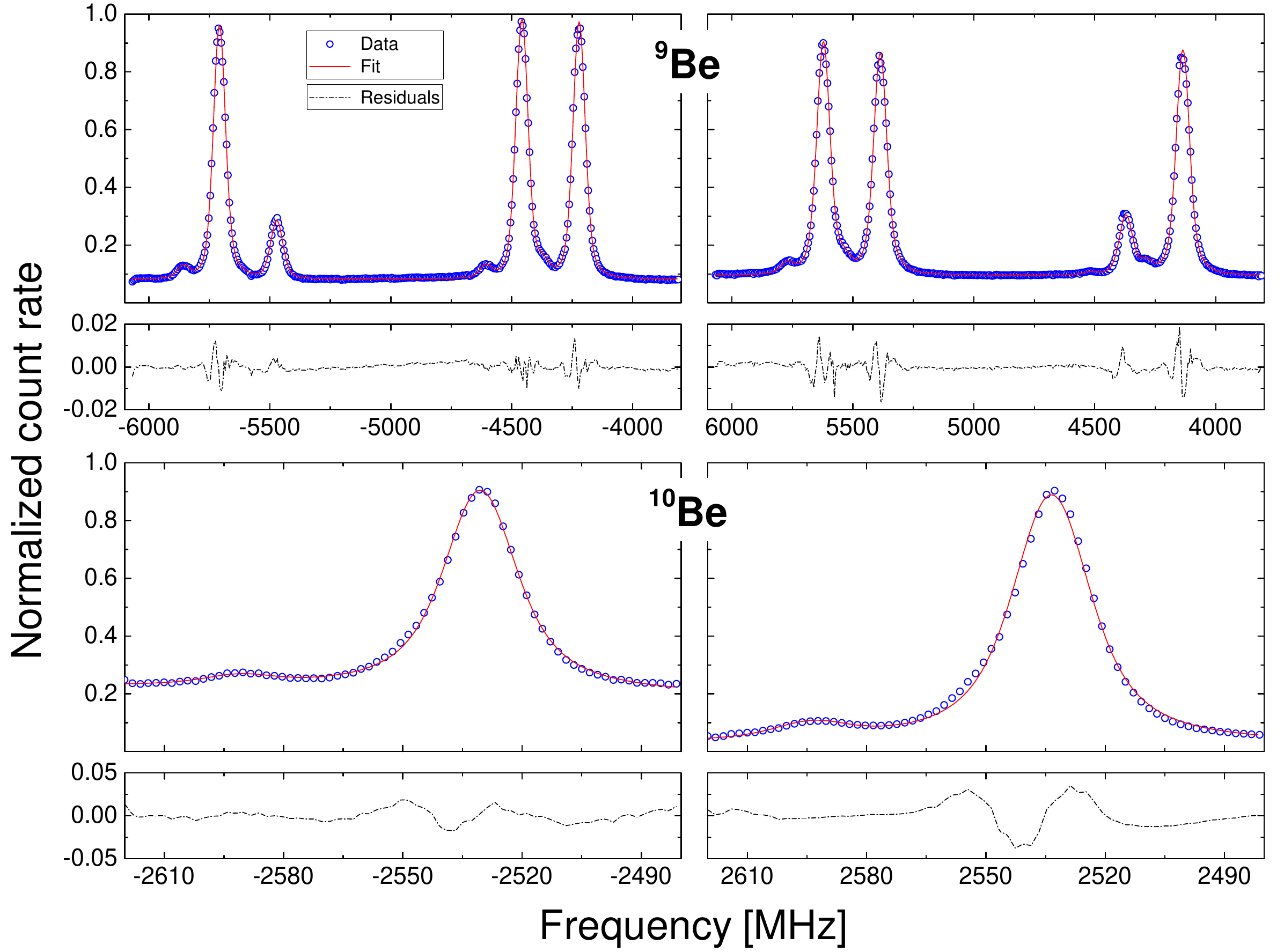}
    \caption{Top: Optical hyperfine spectrum of the \DOne\ transition in 
 $^{9}$Be$^+$ for collinear (left) and anticollinear excitation (right). 
 Spectra were taken at 35-kV ISOLDE voltage and are the sum of 20 
 individual scans with a resolution of 400 channels. Hyperfine spectra 
 are fitted using a multiple Voigt profile for each component (red line, 
 for further details see text). Striking is the appearance of a small 
 satellite peak on the left of each component which is ascribed to energy 
 loss of the ions in inelastic collisions in flight. The differential 
 Doppler-tuning parameter is about 39\,MHz/V. Bottom: Resonance spectra of the \DOne\ transition in $^{10}$Be$^+$ again 
 in collinear and anticollinear geometry also fitted with a Voigt-doublet. A small structure in the residua (shown on the bottom of each graph) remains in all cases, which is discussed in the text.}

 \label{Fig:9benew}
 \end{center}
 \end{figure}

 The position of each hfs sublevel with total angular momentum 
 $\vec{F}=\vec{I}+\vec{J}$, composed of electronic angular momentum $J$ 
 and nuclear spin $I$, and $C=F(F+1)-I(I+1)-J(J+1)$ 
 is given to first order by the hyperfine energy
 \begin{equation}
     \Delta 
 E_{\rm{hfs}}=\frac{A}{2}C+B\cdot\frac{\frac{3}{4}C(C+1)-I(I+1)J(J+1)}{2I(2I-1)J(2J-1)} 
 \ .
      \label{eq:hfsForm}
 \end{equation}
 In the fit function, these shifts determine the spectral line positions 
 relative to the center of gravity. The factors $A$ and $B$ (only for 
 $2p\,^2\mathrm{P}_{\nicefrac{3}{2}}$) of the upper and the lower fine 
 structure state of the transition and the center of gravity 
 $\nu_\mathrm{cg}$ are the free fitting parameters for  the peak 
 positions. The line shape of each component was modelled by two Voigt 
 resonance terms representing the main peak and the satellite peak  
 as discussed above. The 
 distance between the two peaks was fixed to $4$\,V on the voltage axis.
 The Gaussian (Doppler) line width parameter and the intensity ratio 
 between the main peak and the satellite were free parameters but 
 constrained to be identical for all hyperfine components, while the 
 total intensity of each component was also a free parameter. The Lorentzian 
 line width was kept fixed at the natural line width of 19.64\,MHz since 
 significant saturation broadening was not observed.
 Non\-linear least-square minimization of  $\chi^2$ was performed using a 
 Levenberg-Marquardt algorithm.

 Fitting the collinear and the anticollinear spectra independently, we 
 obtain in both cases the centroid frequency $\nu_{\mathrm{cg}}$ of the hyperfine structure.  
However, for calculating the absolute transition rest-frame frequency $\nu_0$ we 
 must take into account that Eq.\,(\ref{eq:vnull})
 requires $\nu_c$ and $\nu_a$ to be measured at the same ion velocity. 
 This is only the case if the center of gravity appears in both spectra 
 at the same voltage. This was accomplished approximately by changing 
 the  frequency of the laser used for anticollinear excitation until the 
 deviation between the corresponding centers of gravity was typically smaller than 
 3\,V, facilitated by the fact that the comb-stabilized laser 
 can be locked at any arbitrary frequency. The remaining small shift 
 $\delta U$ was considered in the analysis by correcting the collinear frequency 
 using the linear approximation $\delta \nu = \nicefrac{\partial 
 \nu}{\partial U} \cdot \delta U$, where $U$ is the total acceleration 
 voltage of the ions which have entered the optical detection 
 region. Hence, the transition rest-frame frequency was calculated 
 according to
 \begin{equation}
      \label{eq:diffdopplertransition}
      \nu_0 = \sqrt{\left(\nu_c - \frac{\partial \nu_{\rm{D}}}{\partial U} 
 \cdot \delta U \right) \cdot \nu_{a}} - \delta \nu_{\rm rec}.
 \end{equation}
 with the differential Doppler shift
 \begin{equation}
     \frac{\partial \nu_{\rm{D}}}{\partial U} = \frac{\nu_0}{mc^2}\left( 
 e+\frac{e(mc^2+eU)}{\sqrt{eU(2mc^2+eU)}} \right)
 \end{equation}
   and the recoil correction term
   \begin{equation}
   \delta \nu_{\rm{rec}} = \frac{h \nu^2_{\rm{photon}}}{m_0c^2}.
   \end{equation}
 The latter takes energy and momentum conservation during the 
 absorption/emission process into account. It contributes with about 
 200\,kHz to the absolute transition frequency and is slightly isotope-dependent.
Each measurement of $\nu_0$ was repeated at least five times for each isotope. Statistical fitting uncertainty of the center of gravity was usually less than 100\,kHz. For each pair of collinear/anti\-collinear spectra the absolute transition frequency was calculated and the final statistical uncertainty was then derived as the standard error of the mean of all measurements being usually of the order of 100--500\,kHz.

 \subsection{Investigations of Systematic Uncertainties}
 \label{sec:systematicuncertainty}

 Sources of systematic errors were investigated on-line in Run I and Run 
 II as well as in an additional test run, when only a previously 
 irradiated target was used to extract the long-lived isotope $^{10}$Be.
 For each isotope about 3--4 different iodine hyperfine transitions were 
 used as reference points for the collinear laser frequency, which 
 implies different locking frequencies of the comb-locked anticollinear 
 laser as well as different offset voltages at the fluorescence detection 
 region. It should be noted that the actual locking frequency of the 
 iodine-locked laser was regularly checked with the frequency comb during 
 each block of measurements.
 
In Run I, the Rb reference clock for the frequency comb was not long-term stabilized on the 1-pps signal and contributed with about 350\,kHz to the systematic uncertainty of the absolute transition frequency \cite{Noe09}.
 
 Additional uncertainties related to 
 the applied acceleration voltages could only arise from the center-of-gravity 
 correction according to Eq.\,(\ref{eq:diffdopplertransition}), which was 
 typically less than 3\,V. The HV-amplification 
 factor, calibrated regularly to better than 
 $3 \cdot 10^{-4}$, leads to uncertainties clearly below the 3-mV level corresponding to approximately 100\,kHz in transition frequency. This contribution can be safely neglected compared to other systematic 
 uncertainties discussed below.

 Additionally, ion and laser beam properties were modified on purpose for 
 investigating a possible influence on the measured transition frequencies 
 and isotope shifts. In deviation from a parallel collimation the ion beam 
 was focused close to 
 the fluorescence detection region with the available electrostatic 
 quadrupole lenses. Similarly, additional convex lenses were added into 
 the light path to focus the laser beams inside the beam line. It was 
 found that these modifications merely changed the signal-to-noise ratio 
 but had no significant influence on the determined resonance frequencies.

 \subsubsection{Laser-Ion-Beam Alignment} \label{sec:alignment}
 The parallel and antiparallel alignment of the respective laser beams 
 with the ion beam was ensured using two apertures inside the beam line. 
 Hence, the range of a possible angle misalignment between laser and ion 
 beam was estimated taking the full aperture of 5\,mm and their distance 
 to each other of 2\,m into account. A conservative estimate with beam 
 diameters of about 4\,mm results in an angle of
 $\alpha = \arctan\left(\Delta z / \Delta x\right) \approx 
 1$~mrad. However, during the preparation of the experiment, both laser 
 beams were superimposed 2~m after the collinear and anticollinear exit 
 windows, respectively. Two extreme cases are to be discussed: The 
 Doppler-shifted frequencies $\nu_{c,a}$ get angle-dependent if both 
 laser beams are well superposed, but are misaligned relative to the 
 ion beam
 \begin{equation}
      \label{eq:dopplerangle}
      \nu_{c,a} = \nu_0 \gamma ( 1 \pm \beta \cdot \cos\alpha).
 \end{equation}
 Then the transition rest-frame frequency $\nu_{0}$ 
 becomes angle-dependent as well
 \begin{eqnarray}
      \label{eq:dopplerangledoble}
      \nu_{0} & = & \frac{1}{\gamma} \sqrt{\frac{\nu_c \cdot \nu_{ac}}{1 
 - \beta^2 \cdot \cos^2\alpha}} \\
      &=& \sqrt{\frac{1 - \beta^2}{1 - \beta^2 \cos^2 \alpha}} 
 \sqrt{\nu_c \cdot \nu_{ac}} \\
       &\approx& \left( 1 - \beta^2 \alpha^2 \right)\sqrt{\nu_c \cdot 
 \nu_{ac}}.
 \end{eqnarray}
Even though the Doppler shift is reduced for both beams, the 
 collinear-anticollinear geometry almost leads to a cancellation of the 
 effect,  the anticollinear resonance is less blue shifted while the 
 collinear resonance is less red shifted. For an angle misalignment of 1 mrad, $\nu_a$ and $\nu_c$ will each be shifted by as much as 1.4\,MHz, whereas the effect on $\nu_0$ is only of the order of about 1\,kHz.

 The other extreme is a misalignment of one laser relative to a well 
 superposed laser-ion-beam pair. Then the transition frequency $\nu_0$ becomes
 \begin{eqnarray}
      \label{eq:dopplerangledoble2}
      \nu_{0} &=& \frac{1}{\gamma} \sqrt{\frac{\nu_c \cdot \nu_{ac}}{(1 - 
 \beta \cdot \cos\alpha) ( 1+\beta)}} \vspace{3mm} \\
      &=& \sqrt{\frac{ 1 - \beta}{(1 - \beta 
 \cos\alpha})}\sqrt{\nu_c \cdot \nu_{ac}} \vspace{3mm}\\
       &\approx& \left( 1 - \beta \alpha^2 \right)\sqrt{\nu_c \cdot 
 \nu_{ac}}.
 \end{eqnarray}
In contrast to Eq.\,(\ref{eq:dopplerangledoble}) this angle-dependence  
can lead under unfavourable conditions to an appreciable shift. 
The influence of the laser-ion-beam alignment was extensively 
studied using a stable $^9$Be ion beam by misaligning one of the 
 laser beams so that a deviation was clearly visible in the horizontal or
 vertical direction. With the typical beam diameter a deviation of about 2\,mm across a distance of $\approx 8$\,m ($\alpha=0.25$\,mrad) was detectable, corresponding to a total effect of about 600\,kHz. Misalignment and realignments where repeated 
 several times but the results of the measurements with misalignment 
 scattered similarly as the measurements with optimized alignment and in both 
 cases the scatter was in accordance with the standard deviation of all 
 regular $^9$Be measurements. During the experiment, the counterpropagating alignment of 
 the laser beams was inspected visually several times per day. A systematic uncertainty of 300\,kHz, corresponding to half the full scattering amplitude was conservatively estimated.

 \subsubsection{Photon Recoil Shift}
 \label{sec:prs}
 \begin{figure*}
 \begin{center}
 {\includegraphics[width=0.98\linewidth,clip=true, trim=5mm 90mm 0mm 0mm] 
 {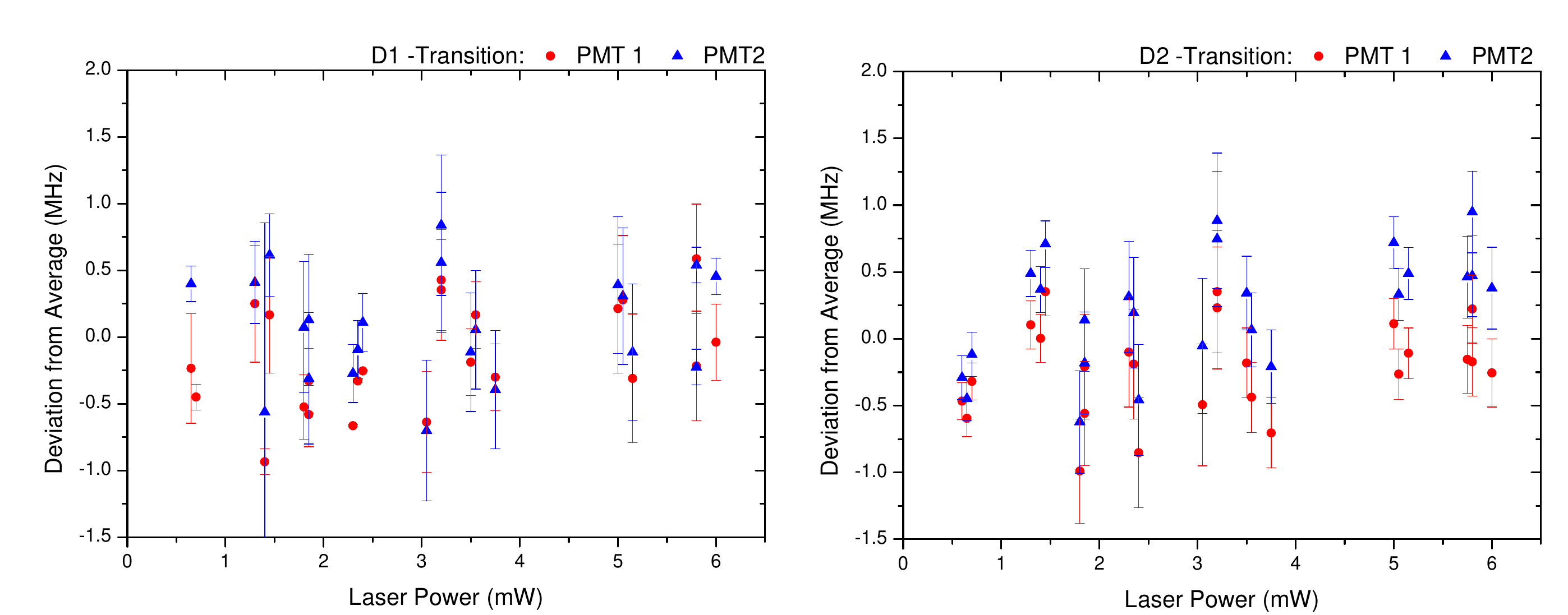}}
    \caption{
    Power dependence of the extracted frequency of 
 $^{10}$Be$^+$ in the D1 (left) and the D2 transition (right). Plotted 
 are the differences to the total mean frequency as a function of 
 laser power. Copropagating and counterpropagating laser beams were 
 adjusted to approximately equal power. Measurements were performed in 
 three series varying the power from the highest to the lowest values. 
 Uncertainties of the individual data points were estimated as the 
 standard deviation for each individual set of three measurements at 
 approximately the same power from this series.
 }
 \label{Fig:bepowerdependence}
 \end{center}
 \end{figure*}
 Repeated interaction with a laser beam can influence the external degrees of motion of an ion or atom as it is well known from laser cooling and 
 laser deceleration in a Zeeman slower. In collinear laser spectroscopy 
 the repeated directed absorption and isotropic re-emission of photons 
 will have the consequence that the ions are either accelerated 
 (collinear excitation) or decelerated (anticollinear excitation). With 
 every absorbed photon, the Doppler-shifted resonance frequency is 
 shifted towards higher frequencies ($\nu_{c}$ and $\nu_{a}$) for both directions and 
 this systematic shift results in a transition frequency 
 $\nu_0$ that is too large. The combination of light ions and 
 ultraviolet photons leads to an exceptionally large photon recoil and 
 the possible influence of this effect must be studied.
 Due to the absence of hyperfine splitting the $2s \rightarrow 2p$ 
 transitions in the even isotopes $^{10,12}$Be$^+$ are closed two-level 
 systems. Hence, the possibility of repeated photon scattering is enhanced 
 compared to the odd-mass isotopes which are pumped into a dark hyperfine state after a few 
 absorption-emission cycles. To investigate whether the photon recoil has 
 a measurable effect, the power dependence of the transition frequency of 
 $^{10}$Be was determined as a function of the laser power.  
 The laser power in both beams was increased stepwise and simultaneously  
 from below 1\,mW up to 6\,mW.
  The deviation of the extracted transition frequencies from the mean 
 frequency determined for the D1- and D2-transition are plotted in 
 Fig.~\ref{Fig:bepowerdependence} as a function of laser power. Each data 
 point is associated with an uncertainty estimated as the standard 
 deviation of a block of three measurements at similar power. In both 
 transitions the peak positions scatter but do not show a common trend 
 upwards or downwards. It appears, however, that we observe at  
 photomultipier tube 2 (PMT2), located about 15\,cm downstream from PMT1, 
 resonances that are systematically higher in frequency than at PMT1. 
 
As a consequence of this observation we have included only data from
 PMT1 in the analysis and have 
 estimated an additional uncertainty for the remaining effect. 
 The average difference between PMT1 and PMT2 is about 300\,kHz and 450\,kHz in the D1 and D2 transition, respectively. Since the distance between the two PMTs is slightly larger than the path of the ions before reaching PMT1, we estimate conservatively a maximal shift of about 400\,kHz for the systematic uncertainty $\Delta \nu_{{\rm{Ph}}}$ caused  by photon recoil.

\subsection{Spectra of the Short-Lived Isotopes $^{7,11,12}$Be}
\label{sec:radioactiveberyllium}

\begin{figure*}
\begin{center}
{\includegraphics[width=\linewidth]{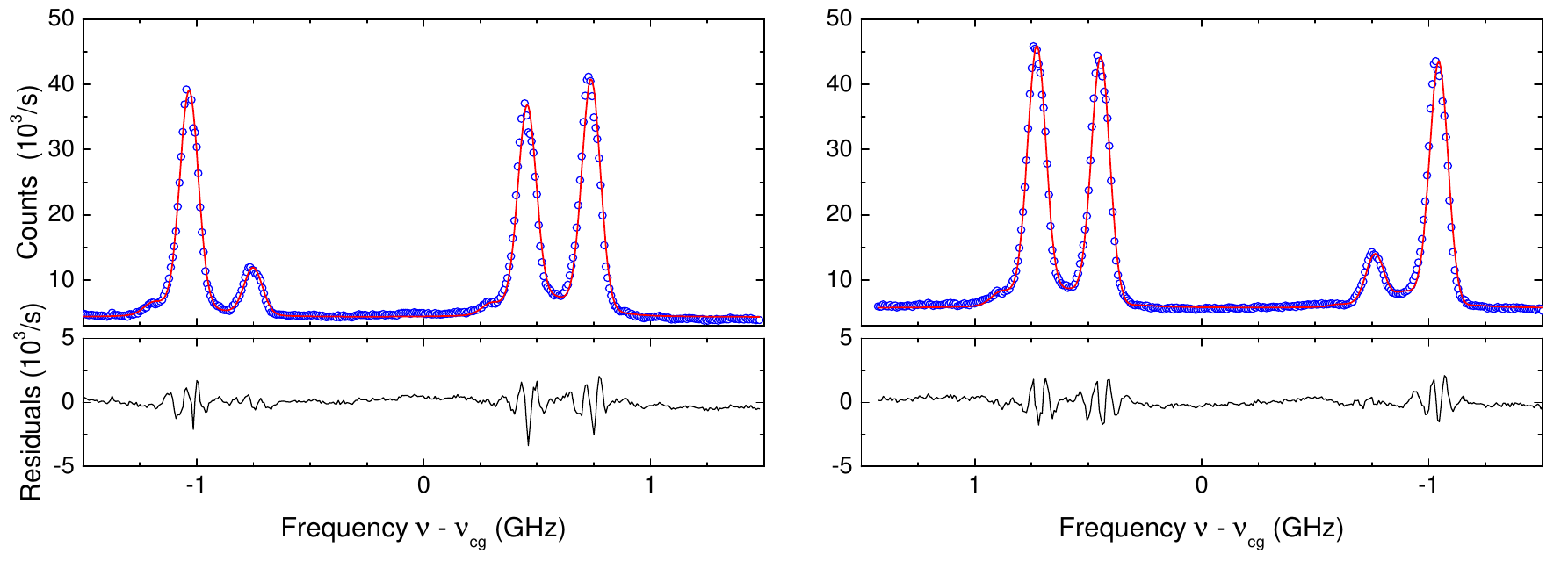}}
   \caption{Spectra obtained in the \DOne transition in $^{7}$Be$^+$ for 
copropagating (left) and counterpropagating laser excitation (right) as 
a sum of 50 individual scans. The data points are fitted with a multiple 
Voigt profile as discussed in the text. In the lower trace the residua 
of the fit are displayed. For more details see \cite{Zak10}.}
\label{Fig:be7}
\end{center}
\end{figure*}

\begin{figure*}
\begin{center}
{\includegraphics[width=\linewidth]{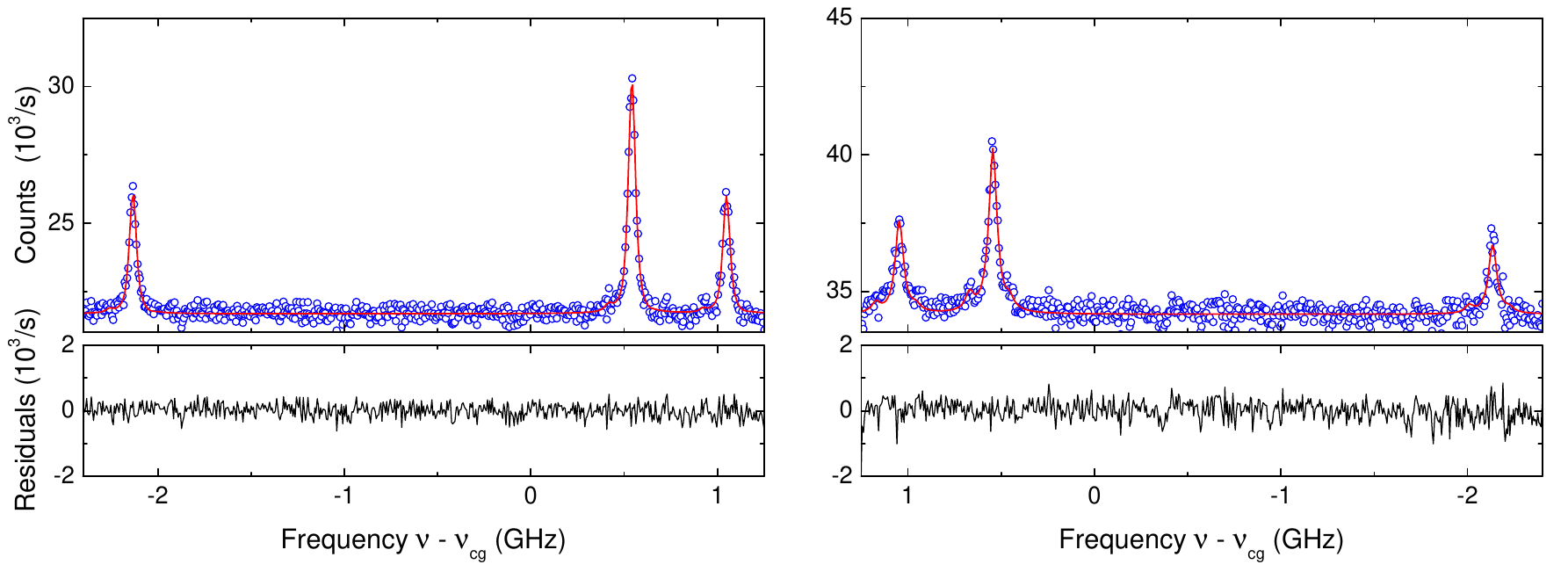}}
   \caption{Resonance spectra of the \DOne transition in $^{11}$Be$^+$ 
in collinear (left) and anticollinear direction (right). The production 
rate was about $10^6$\,ions$/$pulse and thus 400 individual scans were 
accumulated. The data points are fitted with a multiple Voigt profile as 
discussed in the text. Fitting residua are displayed in the lower trace.}
\label{Fig:be11}
\end{center}
\end{figure*}

A typical spectrum of $^{7}$Be is depicted in Fig.\,\ref{Fig:be7}. It is 
the sum of 50 individual scans, taken in the first beam time in 2008. 
Here, line shapes are slightly broader than observed in the second 
beam time \cite{Zak10}. Since the measurements on $^{7}$Be were not repeated in 2010, 
the uncertainties of the fitted line positions are larger 
than for the other isotopes. 
The reduced line width in the second beam time is clearly visible in the 
spectrum of $^{11}$Be depicted in Fig.\,\ref{Fig:be11}. This is the sum of 
400 individual scans.
Spectra of the unresolved hyperfine structure in the \DTwo transitions 
of both these isotopes can be found in \cite{Zak10}.

For the investigation of $^{12}$Be$^+$, proton pulses impinged on the 
target every 3--5 seconds. In this case, spectra in co- and 
counterpropagating geometry were taken by switching the laser beams 
after each proton pulse and photon detection was limited to 60\,ms 
($\approx 3\,T_{\nicefrac{1}{2}}(^{12}\mathrm{Be})$) after the  
pulse to reduce the number of random coincidence events.
\begin{figure*}
\begin{center}
{\includegraphics[width=0.95\textwidth]{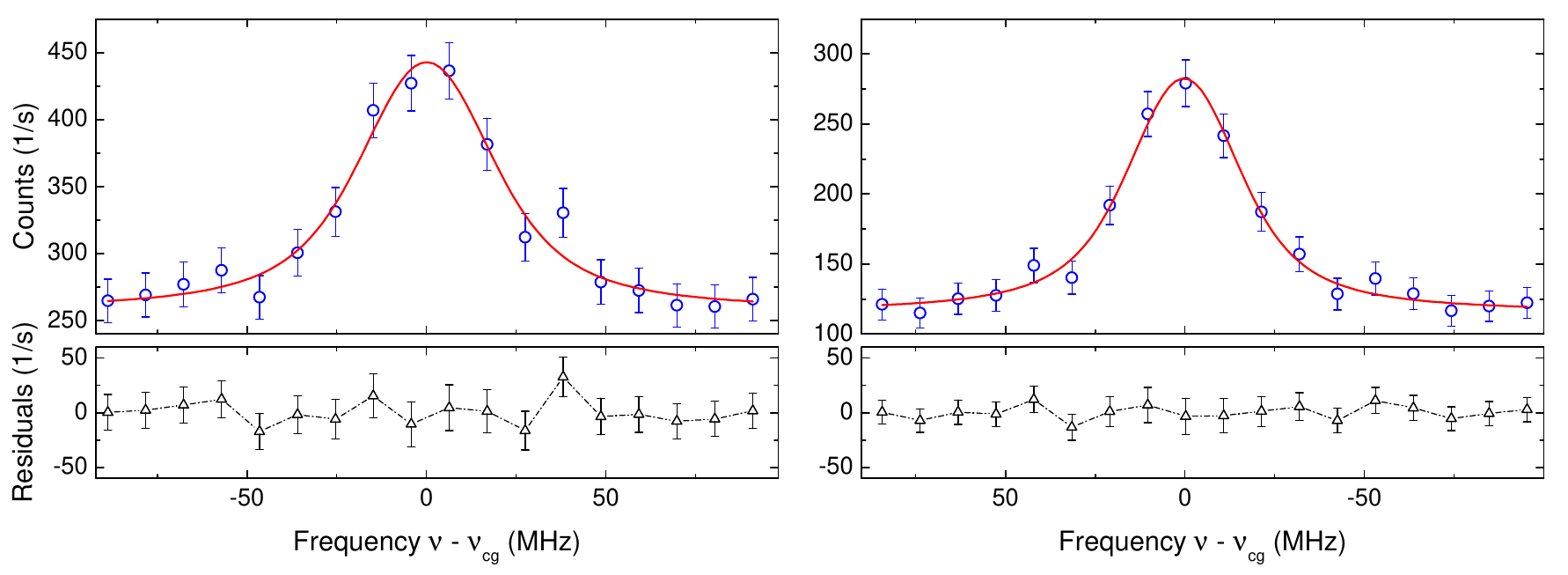}}
   \caption{Resonance spectra of the \DOne transition in $^{12}$Be$^+$ 
for collinear (left) and anticollinear excitation (right) plotted as a 
function of the Doppler-tuning frequency. In total 180 single scans were 
summed up for 8 hours. The symmetric spectra were fitted with a single 
Voigt profile (red line). Residua are displayed in the lower trace.}
\label{Fig:be12coincidence}
\end{center}
\end{figure*}
Figure~\ref{Fig:be12coincidence} shows a typical spectrum which is 
accumulated over 180 individual scans. Here a detection efficiency of about 
1\,photon per 800\,ions was obtained. The Doppler-tuning 
voltage range was restricted to 6\,V corresponding to about 200\,MHz 
frequency span around the main peak, to limit the time required to record 
a single resonance to a few hours. Reference measurements of 
$^{10}$Be$^+$ were interspersed after every 40 single scans to ensure 
stability of all conditions.
Due to the limited statistics not allowing the observation of the small 
satellite peak, only a single Voigt profile was used for fitting the 
resonances. Statistical uncertainty obtained from the fit was usually less than 100\,kHz.

\subsection{Absolute Transition Frequencies}
\label{sec:absolutefrequencies}
\begin{table*}
\begin{center}
\caption{Absolute transition frequencies $\nu_0$ for the \DOne (D1) and 
the \DTwo (D2) transition in  beryllium isotopes obtained in Run I 
(Beamtime 2008) and Run II (Beamtime 2010). The first uncertainty 
represents the statistical,  the second one the total uncertainty 
including systematic uncertainties as discussed in 
Sec.\,\ref{sec:systematicuncertainty}. Systematic uncertainties of the 
results of Run I may be reduced  from those published in 
\cite{Noe09,Zak10} due to the information gained in Run II. Final 
results for all isotopes are printed bold. All values are in MHz.}
\label{tab:expfrequency}
\vspace{2mm}
\begin{tabular}{c c r@{.}l r@{.}l}
\hline\hline
Isotope & Run & \multicolumn{2}{c}{$\nu_0$} &\multicolumn{2}{c}{$\nu_0$} \\
& & \multicolumn{2}{c}{D1}      &\multicolumn{2}{c}{D2} \\
\hline
~$^{7}$Be & I  &  \bf 957\,150\,316&\bf2 (0.8) (0.9) & \bf  
957\,347\,374&\bf5  (0.9) (1.1)  \\
\hline
~$^{9}$Be & I  & 957\,199\,552&9 (0.8) (1.0)   & 957\,396\,616&6  
(1.4)(1.5) \\
  & II & 957\,199\,553&40 (0.12)(0.52)  & \multicolumn{2}{c}{--} \\
  & \bf comb. & \bf 957\,199\,553&\bf28 (0.12)(0.52) &\bf  
957\,396\,616&\bf6  (1.4)(1.5) \\

  & \cite{Bol85} & \multicolumn{2}{l}{957\,199\,652 (120)}     &     
\multicolumn{2}{l}{957\,396\,802 (135)}  \\
\hline
$^{10}$Be & I     & 957\,216\,876&9  (1.4)(1.5)  & 957\,413\,943&9 (0.8) 
(1.0)  \\
$^{10}$Be & II & 957\,216\,876&84 (0.42)(0.66)  & 957\,413\,942&17 
(0.44)(0.70) \\
  & \bf comb. & \bf 957\,216\,876&\bf85 (0.42)(0.66) &\bf  
957\,413\,942&\bf74  (0.44)(0.67) \\
\hline
$^{11}$Be    & I     & 957\,231\,118&1  (1.1)(1.2) & 957\,428\,185&2 
(1.5)(1.6) \\
$^{11}$Be    & II & 957\,231\,118&11 (0.10)(0.52)  & 
\multicolumn{2}{c}{--}         \\
  & \bf comb. & \bf 957\,231\,118&\bf11 (0.10)(0.52) &\bf  
957\,428\,185&\bf2  (1.5)(1.6) \\

\hline
$^{12}$Be & II & \bf 957\,242\,944&\bf86 (0.33)(0.61) &\bf 
957\,440\,013&\bf60 (0.28)(0.58)  \\
\hline\hline\\
\end{tabular}
\end{center}
\end{table*}
Each pair of spectra was fitted as discussed in 
Sec.~\ref{sec:lineshapefitting} to determine the centers of gravity and 
to extract the respective rest frame transition frequency $\nu_0$ 
according to Eq.\,(\ref{eq:diffdopplertransition}). For each isotope and beamtime the weighted mean of all measurements was calculated and results are 
listed in Tab.\,\ref{tab:expfrequency}. All absolute frequencies from 
both beam times agree within their 1-$\sigma$ error bars confirming the 
reproducibility of the measurement. There is a small change compared to 
Table\,I of \cite{Noe15}, namely a slightly larger uncertainty for 
$^{10}$Be from Run II in the D2 line due to a transfer error in the 
statistical uncertainty. Statistical uncertainties in Run II are based 
on the standard error of the mean for typically 4--5 measurements for $^{11,12}$Be and 20--30 measurements for the less exotic isotopes $^{9,10}$Be.
Where available, values from both runs are combined weighted with the 
respective uncertainty. Systematic uncertainties from the second run cannot be further reduced.
These absolute transition frequencies can now be used to evaluate 
differential observables, like the isotope shift, the fine structure 
splitting and the splitting isotope shift.

\subsection{Isotope Shifts and Nuclear Charge Radii} \label{sec:chargeradii}
\begin{table*}
\begin{center}
\caption{Compilation of experimental isotope shifts $\delta 
\nu_{IS}^{9,A}$, obtained from the absolute transition frequencies 
$\nu_0$ in Tab.\,\ref{tab:expfrequency}, field shift 
$\delta\nu_{\rm{FS}}$ extracted as the difference to the theoretical 
mass shifts listed in Tab.\,\ref{tab:propIS} \cite{Puc10} and the 
corresponding change in the mean square nuclear charge radius $\delta 
\left\langle r^2 \right\rangle$ according to Eq.\,(\ref{eq:changer}). 
Results from the \DOne (D1) and the \DTwo (D2) transitions are 
compatible and were combined before the absolute charge radius $R_c$ is 
calculated according to Eq.\,(\ref{eq:absoluter}).}
\label{tab:IsotopeShift}
\vspace{2mm}
\begin{tabular}{l c c c c}
\hline\hline
Isotope & $\delta \nu_{\rm{IS}}^{9,A}$ & $\delta \nu_{\rm{FS}}$ & 
$\delta \left\langle r^2 \right\rangle $ & $R_c$ \\
and transition & /MHz & /MHz & /fm$^2$ & /fm \\
\hline
~$^{7}$Be$^+$ D1       & -49~237.1~(1.1)  & -11.4(1.1) & 0.67~(6)  & \\
~$^{7}$Be$^+$ D2      & -49~242.1~(1.8)  & -10.3(1.8) & 0.61(11)  & \\
~$^{7}$Be$^+$ Mean &                  &            & 0.65~(5)  & 
2.646\,(15)\\

~$^{9}$Be$^+$ D1/D2      & 0                & 0 & 0         & 2.519\,(12)\\

$^{10}$Be$^+$ D1      & 17~323.57(84)    & 13.11(84)  & -0.77~(5) & \\
$^{10}$Be$^+$ D2      & 17~326.1~(1.6)   & 13.6~(1.6) & -0.80~(10)& \\
$^{10}$Be$^+$ Mean &                  &            & -0.78~(4) & 
2.360\,(14)\\

$^{11}$Be$^+$    D1      & 31~564.82\,(74)  & ~4.58(74) & -0.27~(4) & \\
$^{11}$Be$^+$    D2      & 31~568.6~\,(2.2) & ~4.4~(2.2) & -0.26(13) & \\
$^{11}$Be$^+$ Mean &                  &            & -0.27~(4) & 
2.465\,(15)\\

$^{12}$Be$^+$ D1      & 43~391.58\,(80)  & ~1.40(82)  & -0.08~(5) & \\
$^{12}$Be$^+$ D2      & 43~397.0~\,(1.6) & ~1.5~(1.6) & -0.09~(10)& \\
$^{12}$Be$^+$ Mean &                  &            & -0.08~(4) & 
2.502\,(15)\\
\hline\hline\\
\end{tabular}\\
\end{center}
\end{table*}

Isotope shifts are easily obtained as difference of the absolute 
transition frequency of the isotope of interest and the reference 
isotope, which in our case is the stable isotope $^9$Be. The field shift 
$\delta \nu_{\rm{FS}}$, also known as the finite nuclear size or nuclear 
volume effect, can then be extracted according to Eq.\,(\ref{eq:IS}). The 
corresponding mass shifts $\delta \nu_{\rm{MS}}^{9,A}$ as well as the 
field shift constant $F^{9,A}$ were theoretically evaluated to an 
accuracy that exceeds the experimental uncertainty by about an order of 
magnitude and are compiled in Tab.~\ref{tab:propIS}. We have been using 
the results from \cite{Puc10,Puc09} to calculate  $\delta 
\nu_{\rm{FS}}$, which is then combined with the reference radius 
$R_{\mathrm{c}} = 2.519(12)$~fm \cite{Jan72} of 
$^9$Be to provide absolute charge radii along the chain using 
Eq.\,(\ref{eq:absoluter}). It should be noted that the uncertainty of $R_{\mathrm{c}}(^9\mathrm{Be})$ is probably underestimated since C2 scattering from the quadrupole distribution has been omitted, which might change the radius by about 3\% \cite{Sic13}. 

\begin{figure}
\begin{center}
{\includegraphics[width=\linewidth,clip=true]{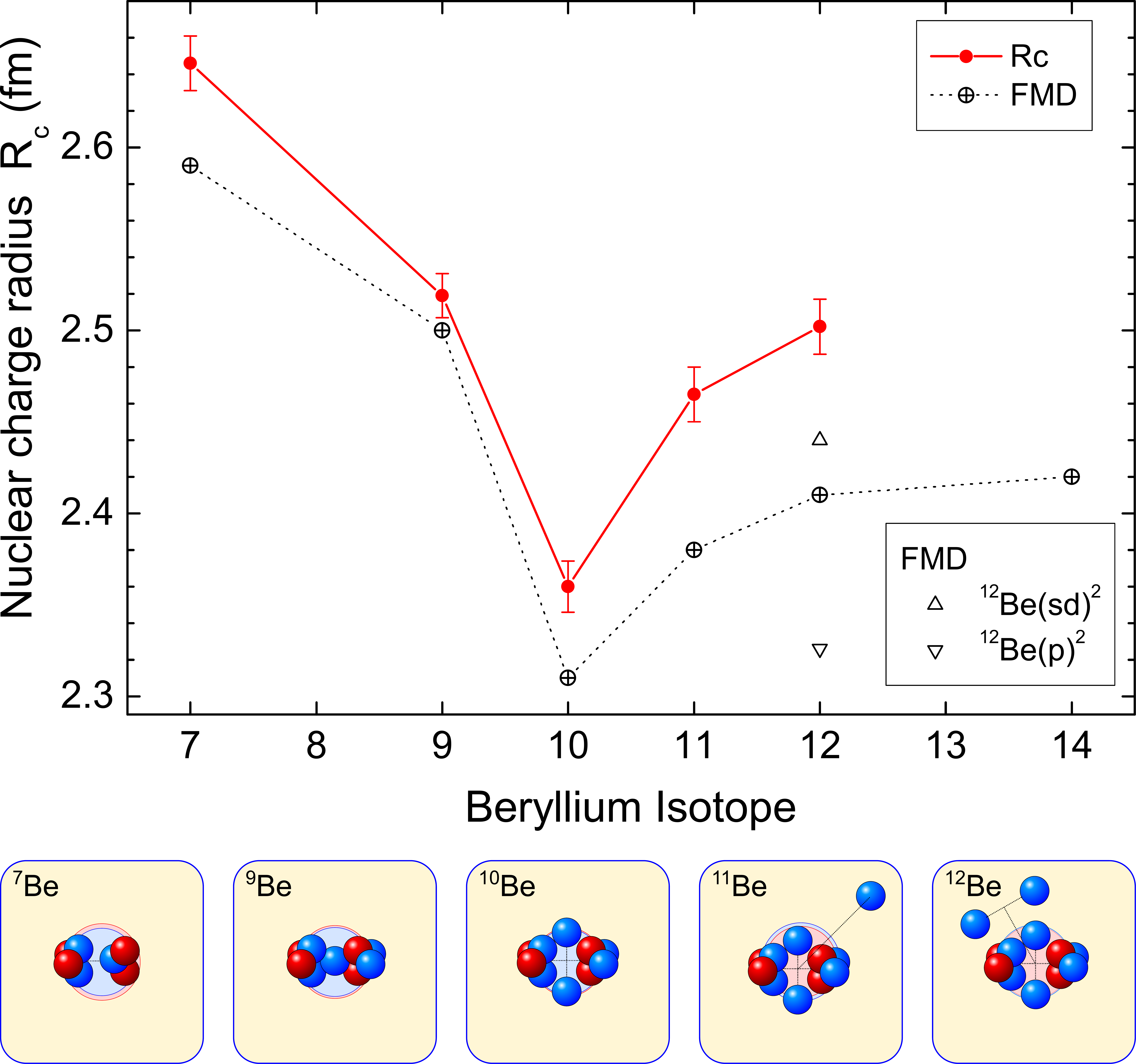}}
   \caption{Nuclear charge radii along the beryllium isotopic chain. The 
reference radius of $^{9}$Be was determined from electron scattering 
experiments \cite{Jan72}. The red bullets ($\bullet$)
   represent the experimental results with uncertainties dominated by 
the uncertainty of the reference charge radius of $^{9}$Be. Therefore 
all uncertainties are similar in size. Additionally shown are results of 
Fermionic Molecular Dynamic (FMD) calculations \cite{Kri12}. For 
$^{10}$Be, calculations were also performed forcing the neutrons into a 
$p^2$ and an $(sd)^2$ orbit. The bottom row shows the structure of the 
isotopes interpreted in a cluster picture. For details see text. }
\label{Fig:be12radii}
\end{center}
\end{figure}

The results are listed in Tab.~\ref{tab:IsotopeShift}. Changes in the 
mean-square charge radii deduced from the isotope shifts in the D1- 
and D2-transition are of similar accuracy in case of even isotopes, 
while for odd-mass isotopes the unresolved hyperfine 
splitting in the D2 lines leads to larger uncertainties. 
Figure~\ref{Fig:be12radii} depicts the development of the rms charge radius along 
the isotopic chain as extracted from the experiment 
($\bullet$) by combining all available data. We have also included 
results from Fermionic Molecular Dynamics (FMD) calculations \cite{Kri12} that follow 
the observed trend quite closely, but the charge radii are generally somewhat 
too small. The two triangles shown for  $^{12}$Be are results of two additional 
calculations  performed under the assumption that the two outermost 
neutrons occupy either a pure
$p^2$ state ($\bigtriangledown$), as expected in the traditional shell 
model or a pure $(sd)^2$ state ($\bigtriangleup$). The latter is 
expected to contribute to the ground state only if the $N=8$ shell gap 
between the $p$-shell and the $sd$-shell is significantly reduced. This 
prediction shows that the charge radius of $^{12}$Be is extremely 
sensitive to the admixture of $sd$-shell states to the ground state, 
which was a strong motivation for the measurement of the $^{12}$Be 
isotope shift.

The trend of the charge radii along the isotopic chain can be understood 
in a simplified picture based on the cluster structure of light 
nuclei \cite{Ash04}. This is visualized in the small panels below the 
graph in Fig.\,\ref{Fig:be12radii}. $^7$Be can be thought of as a 
two-body cluster consisting of an $\alpha$ particle and a helion 
(pnp = $^3$He) nucleus that are bound together and exhibit a considerable 
center-of-mass motion. This motion blurs the proton 
distribution and leads to an increased charge radius.  
$^8$Be is missing since the two $\alpha$ particles  
constituting this nucleus are not bound and the nucleus only exists 
as a resonance. The stable isotope $^9$Be, which has a 
$\alpha+\alpha+n$ structure, is more compact than $^7$Be because the 
$\alpha$ particles themselves are very compact and well bound by the 
additional neutron. This effect is even enhanced with the second neutron 
added in $^{10}$Be. The sudden upward trend to $^{11}$Be is attributed 
to the one-neutron halo character of $^{11}$Be which can be disentangled 
into a $^{10}$Be core and a loosely bound neutron. This halo character 
not only increases the matter radius, but also affects the charge 
radius due to the center-of-mass motion of the core caused by 
the halo neutron. The fact that the charge radius of $^{12}$Be is even 
larger has been related to the fact that the two outermost 
neutrons exhibit a strongly mixed $sd$ character rather than belonging 
to the $p$ shell as expected in the simplified shell-model picture. This 
mixture leads to an increased probability density outside the 
$^{10}$Be core, pulling the $\alpha$ particle apart due to the 
attractive n--$\alpha$ interaction. Theory predicts an $(sd)^2$ admixture of about 
70\% for this nucleus, being a clear 
indication for the disappearance of the classical $N=8$ shell closure. 
For a more detailed discussion of the nuclear charge radii, the 
comparison with ab-initio microscopic nuclear structure calculations 
and the conclusions about the shell closure see 
Refs.\,\cite{Noe09,Kri12,Zak10}.

\subsection{Fine Structure Splitting and Splitting Isotope Shifts} 
\label{sec:FineStructure}

\begin{figure}
\begin{center}
{\includegraphics[width=0.9\linewidth,clip=true]{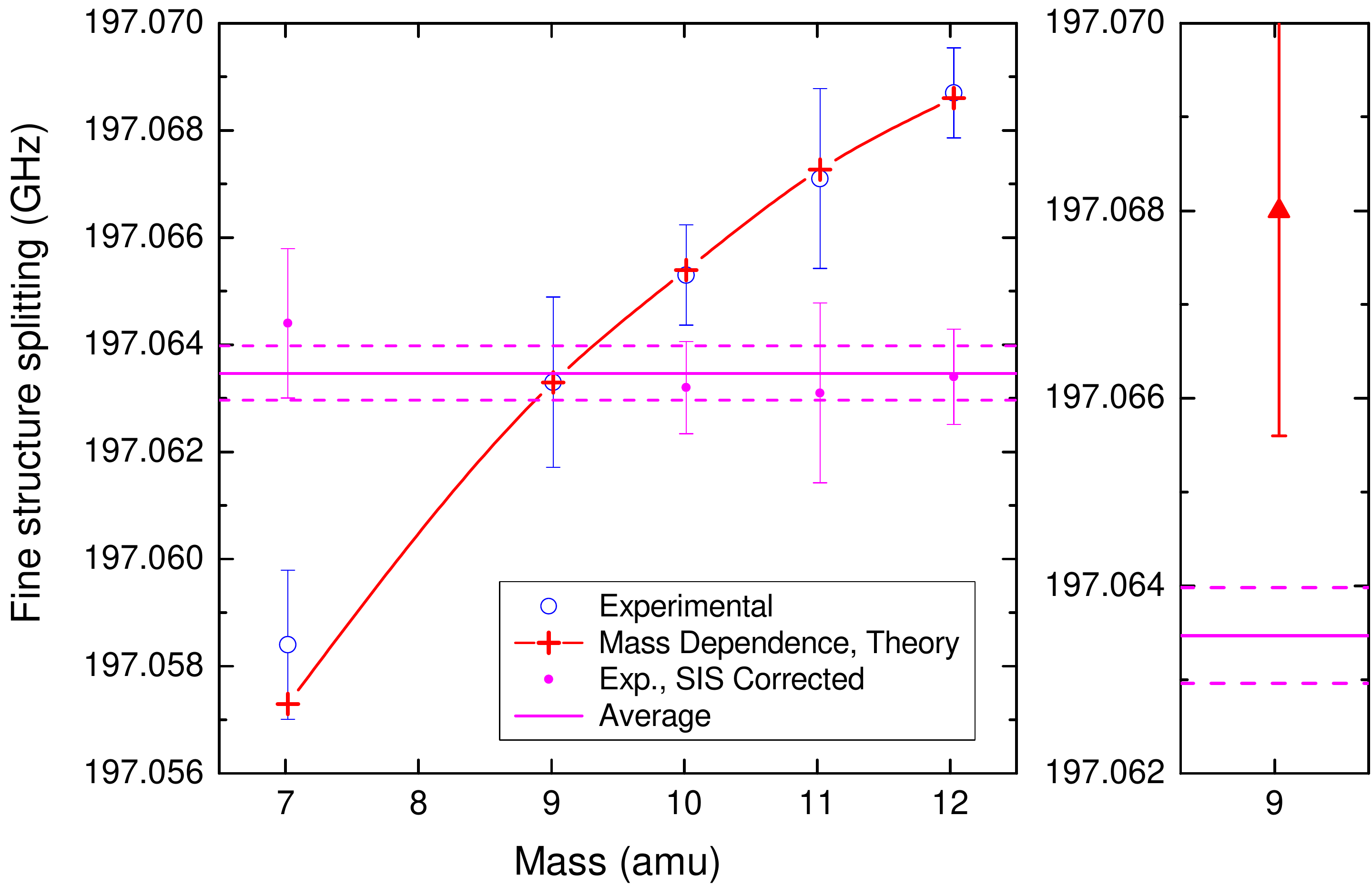}}
   \caption{Left: Mass dependence of the fine structure splitting along 
the berylium isotopic chain. The red curve with crosses shows the 
theoretically expected mass dependence -- the splitting isotope shift 
(sis) -- with respect to $^9$Be. The experimental values (blue circles) 
and the sis-corrected values according to Eq.\,(\ref{eq:fs_projection}) 
(magenta dots) are included. The solid and dotted lines represent the 
mean and standard deviation of the corrected values, respectively. 
Right: Fine structure splitting in $^9$Be from experiment (line) and theory (data point).}
\label{fig:fs_sis}
\end{center}
\end{figure}

\begin{table*}
\begin{center}
\caption{Fine structure splittings $\Delta \nu_{\mathrm{fs}}$, the 
experimental and theoretical \cite{Puc09} splitting isotope shifts 
$\delta \nu_{\mathrm{sis}}$, and the transferred fine structure 
splittings $\Delta \nu_{\mathrm{fs,^ABe\rightarrow\,^9Be}}$ for $^9$Be 
based on the measured splittings in the radioactive isotopes according 
to Eq.\,(\ref{eq:fs_projection}) are listed. The bottom row shows the 
splitting isotope shift between the two even isotopes $^{10}$Be and 
$^{12}$Be. For $\delta \nu^{A,10}_{\mathrm{sis}}$ and $\Delta 
\nu_{\mathrm{fs,^ABe\rightarrow\,^9Be}}$ for $^{7,9,11}$Be after Run II 
required information from Run I since D2 lines of these isotopes were 
not measured in Run II. All values are in MHz.}
\label{tab:fs_splitting}
\vspace{2mm}
\begin{tabular}{c  r@{.}l r@{.}l r@{.}l r@{.}l}
\hline\hline
Isotope & \multicolumn{2}{c}{$\Delta \nu_\mathrm{fs}$} & 
\multicolumn{4}{c}{$\delta \nu^{A,9}_{\mathrm{sis}}$} & 
\multicolumn{2}{c}{$\Delta \nu_{\mathrm{fs,^ABe\rightarrow\,^9Be}}$} \\
   & \multicolumn{2}{c}{D2--D1} & \multicolumn{2}{c}{Exp} & 
\multicolumn{2}{c}{Theory}\\
\hline
~$^{7}$Be &  197\,058&4 (1.4)  & 5&0 (2.1)  & 6&036(1)    & 197\,064&4 
(1.4) \\
~$^{9}$Be &  197\,063&2 (1.6)  &  0&0          & 0&0           & 
197\,063&3 (1.6)  \\
~$^{9}$Be\,$^a$ &\multicolumn{2}{l}{197\,150 (64)} \\

$^{10}$Be & 197\,065&3 (0.9)   & --2&0 (18) & --2&096(1)  & 197\,063&2 
(0.9) \\

$^{11}$Be & 197\,067&1 (1.7)   & --3&8 (23) & --3&965(1)  & 197\,063&1 
(1.7) \\

$^{12}$Be & 197\,068&7 (0.9)   & --5&4 (18)  & --5&300(1)  & 197\,063&4 
(0.8) \\
$^{12-10}$Be & \multicolumn{2}{r}{$\delta 
\nu^{12,10}_{\mathrm{sis}}=$}    & --3&4 (6)       & --3&203 \\
\hline\hline\\
\multicolumn{4}{l}{$^a$ Bollinger \textit{et al.} \cite{Bol85}}\\
\end{tabular}
\end{center}
\end{table*}
From the information provided in Tab.\,\ref{tab:expfrequency} we can 
furthermore extract the fine structure splitting as a function of the 
atomic number. The splitting has previously been measured for $^9$Be with a 
relative uncertainty of about $3 \cdot 10^{-4}$ \cite{Bol85}. The present 
value obtained simply from the difference in $\nu_\mathrm{D1}$ and 
$\nu_\mathrm{D2}$ of this isotope is already 40 times more accurate although its 
accuracy is limited by the unresolved hyperfine structure in the D2 
transition. Only recently the fine structure splitting in 
three-electron atoms became calculable with high precision as  
demonstrated for the case of lithium \cite{Puc14}. Our result confirmed first 
calculations for the $Z=4$ three-electron system of Be$^+$ \cite{Noe15}. 
Moreover, the change in the fine structure splitting along the chain of 
isotopes provides a useful check of the so-called splitting isotope 
shift which can be calculated theoretically to very high accuracy, because the value is
nearly independent of both QED and nuclear volume effects.
This mass dependence is shown in Fig.\,\ref{fig:fs_sis}: the blue data 
points represent the experimental fine structure splittings and the red 
line with red crosses the theoretically expected mass dependence based 
on the calculated splitting isotope shift and the measured splitting 
of $^9$Be. The excellent agreement between the 
theoretical curve and the experimental data can be interpreted as a 
reliable check of the consistency of the mass shift calculations. On the 
other hand, it proves the consistency of the experimental data, because 
the fine structure splittings are based on the combination of absolute 
transition frequencies obtained independently in two beam times, even with 
independent optimization of the experimental conditions. The observation 
that the data points scatter much less than expected from their error 
bars can be ascribed to the fact that our systematic uncertainties largely 
cancel out in considering differential effects. 

The fine structure splitting can most reliably be determined for the 
even-even isotopes $^{10}$Be and $^{12}$Be since there is no hyperfine 
structure which in the D2 transition obscures the 
determination of the center of gravity. This is clearly visible 
in Fig.\,\ref{fig:fs_sis} where the error bars are smallest for these 
two cases. For these isotopes, the fine structure splitting was 
determined sequentially in a short time interval and therefore the best 
cancellation of all systematic uncertainties should occur. Indeed, the 
splitting isotope shift between the two isotopes 
$\delta\nu_\mathrm{sis}^{10,12}=3.43(78)$\,MHz agrees very well with the 
theoretical value of 3.203\,MHz. Here the 
uncertainty of the experimental value is purely statistical assuming 
that all dominant systematic contributions are cancelled out.

With the confidence gained about the reliability of the theoretical 
estimates, we can combine the calculated splitting isotope shifts of 
all isotopes to extract an improved value for the fine structure 
splitting of $^9$Be. To this aim, we correct all measured fine structure 
splittings by  subtracting the theoretical splitting isotope shift
\begin{equation}
\Delta \nu_{\mathrm{fs,^ABe\rightarrow\,^9Be}} = \Delta \nu_{\rm 
fs}(^A{\rm Be}) - \delta \nu^{A,9}_{\mathrm{sis, Theory}}.
\label{eq:fs_projection}
\end{equation}
The results are included in Tab.\,\ref{tab:fs_splitting} and plotted in 
Fig.\,\ref{fig:fs_sis} as magenta bullets. The weighted mean for the  
"isotope-projected" fine structure splitting of $^9$Be represented by the horizontal line is 197\,063.47\,(53)\,MHz. The uncertainty is  
shown by the dashed lines. We believe that a small remaining 
systematic uncertainty of the fine structure splitting is still covered 
by the size of this uncertainty.
In the right hand part of the figure, the theoretical prediction represented 
by the filled red triangle is 
compared with experiment. The calculated splitting in $^9$Be amounts to 
197\,068.0\,(25)\,MHz, which is about 4.5\,MHz larger than the 
experimental value. This difference corresponds to about $1.5\,\sigma$ 
of the combined uncertainties. The theoretical uncertainty is based on 
an estimation of the size of the uncalculated nonlogarithmic terms in 
$m\alpha^7$. These terms are expected to be less than 50\% of the 
calculated leading logarithmic terms.
To further test the QED calculations in lithium-like light systems, similar measurements on ions with higher $Z$, e.g., in boron B$^{2+}$ or carbon C$^{3+}$ are of great interest. At least for B$^{2+}$ the wavelength of about 206\,nm is still achievable and  measurements with the collinear laser spectroscopy technique presented here are planned. 

\section{Summary}
We have demonstrated that quasi-simultaneous collinear-anticollinear laser 
spectroscopy on stable and radioactive isotopes can be used to perform 
precision measurements from which nuclear charge radii even of light and 
very short-lived species can be extracted. At the same time the data can 
be used to perform high-precision tests of fundamental atomic structure 
calculations and bound-state QED. The experimental technique relies on 
the accurate determination of the absolute laser frequency, which has 
become possible with the invention and availability of 
frequency combs. We will apply the technique for further studies: an 
important case is the determination of the charge radius of the 
proton-halo candidate $^8$B for which an experiment is currently under 
preparation at the ATLAS facility at the Argonne National Laboratory.

\begin{acknowledgements}
This work was supported by the Helm\-holtz Association (VH-NG148), the German Ministry for Science and Education (BMBF) under  contracts  05P12RDCIC and 05P15RDCIA, the Helmholtz International Center for FAIR (HIC for FAIR) within the LOEWE program by the State of Hesse, the Max-Planck Society, the European Union 7$^\mathrm{th}$ Framework through ENSAR, and the BriX IAP Research Program No. P6/23 (Belgium). A.~Krieger acknowledges support from the Carl-Zeiss-Stiftung (AZ:21-0563-2.8/197/1). 
\end{acknowledgements}


\begin{thebibliography}{}
%

\bibitem{Noe09} W.~N\"ortersh\"auser {\it et al.}, Phys.\ Rev.\ Lett.\ {\bf 102}, 062503 (2009).

\bibitem{Kri12} A.~Krieger {\it et al.},  Phys.\ Rev.\ Lett.\ {\bf 108}, 142501 (2012).

\bibitem{Noe15}
W.~N\"ortersh\"auser {\it et al.}, Phys.\ Rev.\ Lett.\ {\bf 115}, 033002 (2015).

\bibitem{Neu85} R.~Neugart, Hyp.\ Int.\ {\bf 24}, 159 (1985).

\bibitem{Ott89} E.W.~Otten, \textit{Nuclear Radii and Moments of unstable Isotopes} in: Treatise on heavy ion science Vol. 8, ed. D.A. Bromley, New York: Plenum Publishing Corp. (Springer), 517 (1989).

\bibitem{Bil95} J.~Billowes and P.~Campbell, J.\ Phys.\ G {\bf 21}, 707 (1995).

\bibitem{Neu02} R.~Neugart, Eur.\ Phys.\ J.\ A {\bf 15}, 35 (2002).

\bibitem{Neu06} R.~Neugart and G.~Neyens, Lect.\ Notes Phys.\ {\bf 700}, 135 (2006).

\bibitem{Che10} B.~Cheal and K.~Flanagan, Jour. of Phys. G {\bf 37}, 113101 (2010).

\bibitem{Bla13} K.~Blaum, J.~Dilling and W.~N\"ortersh\"auser, Phys.\ Scr.\ {\bf  T152}, 014017 (2013).

\bibitem{Cam15} P.~Campbell, I.D.~Moore and M.R.~Pearson, Prog.\ Part.\ Nucl.\ Phys.\ {\bf 86}, 127 (2016).

\bibitem{Din91}
T.P.~Dinneen, N.~Berrah-Mansour, H.G.~Berry, L.~Young, and R.C.~Pardo, Phys.\ Rev.\ Lett.\ {\bf 66}, 2859 (1991).

\bibitem{Tho98} J.K.~Thompson, D.J.H.~Howie, and E.G.~Myers, Phys.\ Rev.\ A  {\bf 57}, 180 (1998).

\bibitem{Mye99} E.G.~Myers, H.S.~Margolis, J.K.~Thompson, M.A.~Farmer, J.D.~Silver, and M.R.~Tarbutt, Phys.\ Rev.\ Lett.\  {\bf 82}, 4200 (1999).

\bibitem{Gei99} W.~Geithner {\it et al.}, Phys.\ Rev.\ Lett.\ {\bf 83}, 3792 (1999).

\bibitem{Mar11} K.~Marinova {\it et al.}, Phys.\ Rev.\ C {\bf 84}, 034313 (2011).

\bibitem{Yan00} Z.-C.~Yan and G.W.F.~Drake, Phys.\ Rev.\ A {\bf 61}, 022504 (2000).

\bibitem{Yan03} Z.-C.~Yan and G.W.F.~Drake, Phys.\ Rev.\ Lett.\ {\bf 91}, 113004 (2003).

\bibitem{Puc06} M.~Puchalski, A.M.~Moro and K.~Pachucki, Phys.\ Rev.\ Lett.\ {\bf 97}, 133001 (2006).

\bibitem{Yan08} Z.-C.~Yan, W.~N\"ortersh\"auser and G.W.F.~Drake, Phys.\ Rev.\ Lett.\ {\bf 100}, 243002 (2008).

\bibitem{Yan08b} Z.-C.~Yan, W.~N\"ortersh\"auser, and G.W.F.~Drake,  Phys.\ Rev.\ Lett.\ {\bf 102}, 249903(E) (2009).

\bibitem{Noe11} W.~N\"ortersh\"auser {\it et al.}, Phys.\ Rev.\ A {\bf 83}, 012516 (2011).

\bibitem{Pac04} K.~Pachucki and J.~Komasa, Phys.\ Rev.\ Lett.\ {\bf 92}, 213001 (2004).

\bibitem{Puc13} M.~Puchalski, K.~Pachucki, and J.~Komasa, Phys.\ Rev.\ A {\bf 89}, 012506 (2014).

\bibitem{Puc15b} M.~Puchalski, J.~Komasa, and K.~Pachucki, Phys.\ Rev.\ A {\bf 92}, 062501 (2015).

\bibitem{Tan85} I.~Tanihata {\it et al.}, Phys.\ Rev.\ Lett.\ {\bf 55}, 2676-2679 (1985).

\bibitem{Wan04} L.-B.~Wang {\it et al.},  Phys.\ Rev.\ Lett.\ {\bf 93}, 142501 (2004).

\bibitem{Mue07} P.~M\"uller {\it et al.},  Phys.\ Rev.\ Lett.\ {\bf 99}, 252501 (2007).

\bibitem{Ewa04} G.~Ewald {\it et al.},  Phys.\ Rev.\ Lett.\ {\bf 93}, 113002 (2004).

\bibitem{San06} R.~S\'anchez {\it et al.}, Phys.\ Rev.\ Lett.\ {\bf 96}, 033002 (2006).

\bibitem{Pac10} K.~Pachucki and V.A.~Yerokhin, Phys.\ Rev.\ Lett.\ {\bf 104}, 070403 (2010).

\bibitem{Puc14} M.~Puchalski and K.~Pachucki, Phys.\ Rev.\ Lett.\ {\bf 113}, 073004 (2014).

\bibitem{Nob06} G.A.~Noble, B.E.~Schultz, H.~Ming, and W.A.~van Wijngaarden, Phys.\ Rev.\ A   {\bf 74}, 012502 (2006).

\bibitem{San11} C.J.~Sansonetti, C.E.~Simien, J.D.~Gillaspy, J.N.~Tan, S.M.~Brewer, R.C.~Brown, S.~Wu, and J.V.~Porto, Phys.\ Rev.\ Lett.\ {\bf 107}, 023001 (2011).

\bibitem{Bro13} R.C.~Brown, S.J.~Wu, J.V.~Porto, C.J.~Sansonetti, C.E.~Simien, S.M.~Brewer, J.N.~Tan, and J.D.~Gillaspy, Phys.\ Rev.\ A  {\bf 87}, 032504 (2013).

\bibitem{Puc08} M.~Puchalski and K.~Pachucki, Phys.\ Rev.\ A {\bf 78}, 052511 (2008).

\bibitem{Lu13} Z.T.~Lu, P.~Mueller, G.W.F.~Drake, W.~N\"ortersh\"auser, S.C.~Pieper, and Z.C.~Yan, Rev.\ Mod.\ Phys.\ {\bf 85}, 1383 (2013).

\bibitem{Dra10} G.W.F.~Drake, priv. comm. (2010).

\bibitem{Puc10} M.~Puchalski and K.~Pachucki, Hyp.\ Int.\ {\bf 196}, 35 (2010).

\bibitem{Pac11} K.~Pachucki, priv. comm. (2011).

\bibitem{Jan72} J.A.~Jansen, R.T.~Peerdeman and C.~de~Vries, Nucl.\ Phys.\ A {\bf 188}, 337 (1972).

\bibitem{Der08}  A.~Derevianko, S.G.~Porsev, and K.~Beloy, Phys.\ Rev.\ A {\bf 78}, 010503(R) (2008).

\bibitem{Dou74} M.~Douglas and N.M.~Kroll, Ann.\ Phys.\ (N.Y.) {\bf 82}, 89 (1974).

\bibitem{Puc15} M.~Puchalski and K.~Pachucki, Phys.\ Rev.\ A {\bf 92}, 012513 (2015).

\bibitem{Yan02} Z.~C.~Yan {\it et. al.}, Phys.\ Rev.\ A, {\bf 66}, 042504, 1-8, (2002).

\bibitem{Fed08} V.~N.~Fedosseev {\it et al.},  Nucl.\ Instrum.\ Meth.\ B {\bf 266}, 4378 (2008).

\bibitem{Koe98} U.~Koester {\it et al.}, Proc.\ of ENAM98: Exotic Nuclei and Atomic Masses, Bellaire, Michigan, USA, Am.\ Inst.\ Phys.\ Conf.\ Proc.\ {\bf 455}, 989 (1998).

\bibitem{Iso13} $\mathrm{https://oraweb.cern.ch/pls/isolde/query\_tgt}$, ISOLDE database (2016).

\bibitem{Oka08} K.~Okada {\it et al.}, Phys.\ Rev.\ Lett.\ {\bf 101}, 212502 (2008).

\bibitem{Win83} D.J.~Wineland, J.J.~Bollinger, and W.M.~Itano, Phys.\ Rev.\ Lett.\ {\bf 50}, 628 (1983).

\bibitem{Fia92} D.~C.~Fiander {\it et al.},CERN/PS92-38, Proc.\ of 20th Power Modulator Symposium (1992)

\bibitem{Aud97} G.~Audi, O.~Bersillon, J.~Blachot, and A.H.~Wapstra, Nucl.\ Phys.\ A {\bf 624}, 1 (1997).

\bibitem{Neu81} R.~Neugart, Nucl.\ Instr.\ Meth.\ {\bf 186}, 165 (1981).

\bibitem{Buc82} F.~Buchinger {\it et al.}, Nucl.\ Instr.\ Meth.\ B {\bf 202}, 159 (1982).

\bibitem{Mue83} A.~C.~M\"uller {\it et al.}, Nucl.\ Phys.\ A {\bf 403}, 234 (1983).

\bibitem{Neu86} R.~Neugart {\it et al.},  Nucl.\ Instr.\ Meth.\ B {\bf 17}, 354 (1986).

\bibitem{Gei05} W.~Geithner {\it et al.}, Phys.\ Rev.\ C {\bf 71}, 064319 (2005).

\bibitem{Ney05} G.~Neyens {\it et al.}, Phys.\ Rev.\ Lett.\ {\bf 94}, 022501 (2005).

\bibitem{Kow05} M.~Kowalska {\it et al.}, Eur.\ Phys.\ J.\ A {\bf 25}s01, 193 (2005).

\bibitem{Thu09} T. Th\"ummler {\it et al.}, New J.\ Phys.\ {\bf 11}, 103007  (2009).

\bibitem{Kri11} A.~Krieger {\it et al.}, Nucl.\ Instr.\ Meth.\ A {\bf 632}, 23 (2011).

\bibitem{Pou88} O.~Poulsen and E.~Riis, Metrologia {\bf 25}, 147 (1988).

\bibitem{Rii94} E.~Riis, A.G. Sinclair, O.~Poulsen, G.W.F.~Drake, W.R.C.~Rowley, and A.P.~Levick, Phys.\ Rev.\ A {\bf 49}, 207 (1994).  

\bibitem{Kno04} H.~Knoeckel {\it et al.}, Toptica Photonics (2004).

\bibitem{Bol85} J.J.~Bollinger,J.S.~Wells, D.J.~Wineland, and W.M.~Itano, Phys.\ Rev.\ A {\bf 31}, 2711 (1985).

\bibitem{Zak10} M.~Zakova {\it et al.}, J.\ Phys.\ G {\bf 37}, 055107 (2010).

\bibitem{Puc09} M.~Puchalski and K.~Pachucki, Phys.\ Rev.\ A {\bf 79}, 032510 (2009).

\bibitem{Sic13} I. Sick, Universit\"at Basel, priv. comm. (2013). 

\bibitem{Ash04} N.I.~Ashwood, Phys.\ Lett.\ B {\bf 580}, 129 (2004).


\end{thebibliography}
{}

\end{document}